\documentclass[prd,aps]{revtex4}
\usepackage{amssymb,amsmath,amsfonts,amsbsy,epsfig}
\usepackage{slashed}
\usepackage{graphicx}
\usepackage{physics}
\usepackage[utf8]{inputenc}
\usepackage[caption=false]{subfig}
\usepackage{graphicx}

\usepackage{hyperref}
\usepackage{physics}
\usepackage{adjustbox}

\newcommand{\aj}{AJ}
\newcommand{\mnras}{MNRAS}  

\begin{document}

\title{A nontrivial footprint of standard cosmology in the future observations of low--frequency gravitational waves}
\date{\today}

\author{
Jorge Alfaro$^*$ and Mauricio Gamonal$^\dagger$
}

\affiliation{
Instituto de F\'isica, Pontificia Universidad Cat\'olica de Chile, Av. Vicu\~na Mackenna 4860, Santiago, Chile.\\
$^*$jalfaro@fis.puc.cl, $^\dagger$mfgamonal@uc.cl
}

\begin{abstract} 
\noindent Recent research show that the cosmological components of the Universe should influence on the propagation of Gravitational Waves (GWs) and even it has been proposed a new way to measure the cosmological constant using Pulsar Timing Arrays (PTAs). However, these results have considered very particular cases (e.g. a de Sitter Universe or a mixing with non-relativistic matter). In this work we propose an extension of these results, using the Hubble constant as the natural parameter that includes all the cosmological information and studying its effect on the propagation of GWs. Using linearized gravity we considered a mixture of perfect fluids permeating the spacetime and studied the propagation of GWs within the context of the $\Lambda$CDM model. We found from numerical simulations that the timing residual of local pulsars should present a distinguishable peak depending on the local value of the Hubble constant. As a consequence, when assuming the standard $\Lambda$CDM model, our result predicts that the region of maximum timing residual is determined by the redshift of the source. This framework represents an alternative test for the standard cosmological model, and it can be used to facilitate the measurements of gravitational wave by ongoing PTAs projects.
\end{abstract}

\maketitle 

\section{Introduction}
\label{sec:intro}

In the last two decades a lot of astrophysical evidence has been found suggesting that our Universe is expanding at an accelerated rate \citep{Perlmutter1996,Riess1998,Planck2018,Riess2018,LIGO2019_GW,Holicow2019}. This observation is the cornerstone of modern cosmology and represents an ideal setting in which the large scale aspects of gravitation can be tested. A century after the formulation of General Relativity \citep{Einstein1916}, it remains as the most successful description of gravity, and after many attempts, the Lambda--Cold Dark Matter ($\Lambda$CDM) model has become the standard theoretical framework in which we can study the cosmological phenomena, describing our flat, isotropic, homogeneous and expanding Universe.\\

Within the $\Lambda$CDM model, the main quantity, among others, that describes the rate of expansion is the Hubble parameter and its value at the present day, the \textit{Hubble constant}, is denoted by $H_0$. The value of this constant represents the current rate of expansion and contains information of the composition of the Universe. Recently there has been great controversy about its actual value, in particular, from the tension in the data obtained from the early \citep{Planck2018} and the late Universe \citep{Riess2018,Holicow2019}. This fact has called the local behavior of $H_0$ into question, and therewith many potential explanations to the phenomenon have been raised: From possible new physics to as-yet unrecognized uncertainties from the observations \citep{Odderskov2014,Ko2016,Freedman2017,Bringmann2018,Camarena2018,Mortsell2018,DiValentino2018,Feeney2019}. Nevertheless, no consensus has been reached so far and an intense debate is still going on, waiting for improved empirical data or more robust and complete theories of gravitation. \\

On the other hand, the measurement of \textit{Gravitational Waves} (GWs) carried out by the ground-based detector LIGO \citep{Ligo2016}, is considered as one of the last experimental verifications of general relativity. Actually, GWs have been used as a new alternative for the measurement of cosmological parameters: A gravitational--wave standard siren was used to measure the Hubble constant independently \citep{Ligo2017}. This kind of observations show us that the efforts involved in the accurate measurement of gravitational waves could be useful in the analysis of cosmological parameters within the next years. Nevertheless, ground-based detectors are not the unique way to detect GWs. Another type of ongoing gravitational-wave experiment is the \textit{Pulsar Timing Array} (PTA), which uses the residual time of the arriving electromagnetic emissions from different millisecond-pulsars located in the Milky Way and their correlations to determine the presence of gravitational radiation \citep{Barke2015,Hobbs2017,Nano2019_fp,Burke2019}. Many projects have been developed, as NANOGrav \citep{NanoGrav2013}, Parkes \citep{Parkes2013} or the European PTA \citep{EPTA2010}, members of the International Pulsar Timing Array collaboration \citep{Hobbs2010} (IPTA), which together, seek to identify and measure low-frequency (i.e. $\sim10^{-9}$ to $10^{-8}$ Hz) gravitational waves coming from astrophysical sources as an isotropic and stochastic background \citep{EPTA_BG,IPTA2016,NANOGrav} or Continuous Gravitational Waves (CGWs) from individual Supermassive Binary Black Holes (SMBBHs) \citep{Babak2016,Mingarelli2017,Nano2019_mm}, in which PTA experiments can detect timing differences of $\sim 100$ ns and the dimensionless strain amplitude of CGWs is expected to be $h \sim 10^{-15}$ for sources at redshift $z\lesssim0.5$. The second data release of the IPTA collaboration was published recently \cite{IPTA_2DR}, but the results are not conclusive yet \\

Considering the previous facts, the work of \citet{Bernabeu:2011if} shows the effects produced by a non--zero cosmological constant $\Lambda$ on the propagation of gravitational waves, which includes corrections of order $\sim \sqrt{\Lambda}$ to the phase and the amplitude. Later, it was found \cite{Espriu:2012be,Espriu:2014mwa} that certain modifications to the frequency (i.e. the usual redshift) and a non--trivial correction to the wave number can be found. Furthermore, in those papers was proposed that the magnitude of residual time in a PTA experiment could change due to the action of the cosmological constant. A wide explanation and a review of this phenomenon can be found in the work of \citet{Alfaro:2017sxd}, where the action of non-relativistic matter was included into the phenomenon, showing that Dark Matter increases the effect of $\Lambda$ on the propagation of gravitational waves. In that sense, we expect that each cosmological component of the Universe affects the propagation of GWs similar to the case of $\Lambda$, reason of why we will follow the idea of using the Hubble constant as the main control parameter.\\

The manuscript is structured as follows: In Section \ref{sec:Gravitational Waves in an expanding Universe} we develop the theoretical formalism of our framework, using the linearization of General Relativity and the $\Lambda$CDM model to show how the Hubble constant affects the propagation of Gravitational Waves. The entire Section \ref{sec:Using PTA to measure the Hubble constant} is devoted to analyze the possibility of using PTA experiments to observe the studied phenomenon: In subsection \ref{subsec:Timing residual and the working of PTA} we shortly explain the working of PTA and in subsection \ref{subsec:Including the LCDM model and numerical analysis} we show how our framework can be tested using the previous setup. Furthermore, in subsection \ref{subsec:A relationship between PTA observables and the Hubble constant} we present a reasonable approximation that give us a relationship between the location of the maximum timing residual and the redshift of the source. Finally, in Section \ref{sec:Conclusions} we present our conclusions and an outlook for future research.

\section{Gravitational Waves in an expanding Universe}
\label{sec:Gravitational Waves in an expanding Universe}
\subsection{Linearized Theory of General Relativity and Standard Cosmology}
\label{subsec:Linearized Theory of General Relativity and Standard Cosmology}

The classical theory of General Relativity, finally consolidated by \citet{Einstein1916}, predicts that matter and the curvature of spacetime are related through the Field Equations,
\begin{equation}
\label{EFE}
G_{\mu\nu} + \Lambda g_{\mu\nu} = \kappa T_{\mu\nu},
\end{equation}
where $\kappa=8\pi G/c^4$, $G_{\mu\nu}$ are the components of the Einstein tensor, $T_{\mu\nu}$ are the components of the stress-energy tensor and $\Lambda$ is the cosmological constant, which plays an important role within the context of cosmology since it can be interpreted as the energy related to the vacuum, called commonly \textit{Dark Energy}. The prediction of the existence of gravitational radiation \citep{Einstein1916_GW}, came from the linearization of \eqref{EFE}, which is nothing else than doing perturbation theory around a flat spacetime,
\begin{equation}
g_{\mu \nu} = \eta_{\mu \nu} + h_{\mu \nu}, \qquad\abs{h_{\mu\nu}} \ll 1.
\end{equation}  

The computation of the linearized Field Equations can be found in any text of General Relativity \citep{cheng}, giving the following expression,
\begin{equation}
\label{linearized_EFE}
\square \bar{h}_{\mu \nu} = -2 \Lambda \eta_{\mu \nu}- 2 \kappa T_{\mu \nu},
\end{equation}
where $\bar{h}_{\mu\nu}$ is the trace-reversed perturbation defined by 
\begin{equation}
\bar{h}_{\mu \nu} \equiv h_{\mu \nu} - \frac{1}{2} \eta_{\mu \nu} h,
\end{equation}
which satisfies the Lorenz gauge condition 
\begin{equation}
\partial_\beta \bar{h}^{\beta \alpha} = 0.
\end{equation}

Note that there is still gauge freedom, but we can completely fix it using the Transverse--Traceless gauge, in which $\bar{h}_{\mu\nu} = h_{\mu\nu}$. Therefore, the metric perturbation can be decomposed \citep{Alfaro:2017sxd} as $h_{\mu\nu} = h_{\mu\nu}^{(\textrm{GW})} + h_{\mu\nu}^{(\Lambda)}+ h_{\mu\nu}^{(\textrm{bg})}$, which satisfy the following field equations,

\begin{equation}
\label{decomposition}
\square {h}_{\mu\nu}^{(GW)} = 0,\quad\square {h}_{\mu\nu}^{(\Lambda)} = - 2\Lambda \eta_{\mu\nu},\quad
{h}_{\mu\nu}^{(\textrm{bg})} = -2 \kappa T_{\mu\nu}.
\end{equation}

The linearization is well justified because the perturbations from the Minkowski spacetime are very small when we are far away from the source, which will be our case of study. As explained by \citet{Bernabeu:2011if} and \citet{Espriu:2014mwa}, by using a time--independent --and at order $\sqrt{\Lambda}$-- coordinate transformation, it can be shown that the solution of $h_{\mu\nu}^{(\Lambda)}$ corresponds to the linearization of a static spherically symmetric metric (i.e. the Schwarzschild-de Sitter metric) in the Lorenz gauge. In this work the relevant corrections will be of order $H_0$ (i.e. $\sim \sqrt{\Lambda}$), and therefore, the analysis of GWs and the decomposition \eqref{decomposition} will be simpler in the Lorenz gauge. As the contribution of different terms turns to be additive, the discussion seems straightforward. \\

If the source is located at a large (but non cosmological) distance, the perturbation will be described approximately by a combination of harmonic functions \citep{Espriu:2014mwa,Alfaro:2017sxd}
\begin{equation}
h^{\textrm{(GW)}}_{\mu \nu} = \frac{A_{\mu\nu}}{r}  \sin[k_\mu x^\mu]  \label{GW_solution_harmonic} + \order{H_0^2},
\end{equation}
where $A_{\mu\nu}$ are the components of the polarization tensor, $k^\mu$ is the 4--wavevector and $x^\mu$ is the 4--position. We will denote $\Omega$ as the angular frequency of a \textit{monochromatic} gravitational wave. It is important to note that the coordinates $\{t,r\}$ corresponds to the spherically symmetric context explained before and they will represent the coordinates which origin is placed at the --usually spherical shaped-- source of GWs.\\  

On the other hand, it is commonly accepted that we live in a mostly isotropic and homogeneous Universe. As a first approximation, we can use the FLRW metric to describe the geometry of spacetime \citep{Cervantes2011}, which is of the form
\begin{equation}
\label{FLRW_metric}
\dd s^2 = - \dd T^2 + a^2(T) \left( \frac{\dd R^2}{1-k R^2} + R^2 \dd \Omega^2 \right),
\end{equation}
where $a(T)$ is the adimensional scale factor and the spacetime is described by comoving coordinates $\{T,R\}$. In this work we will use $k=0$, which represents a globally flat geometry of the Universe, as it is currently observed \citep{Planck2018}.  If we consider different perfect fluids as material components of the Universe, and label them with the subscript $i$, having each of them an energy density $\rho_i$, isotropic pressure $p_i$ and a equation of state $p_i=\omega_i \rho_i$, the Friedmann equations will give us the following expression for the energy density of the i--th fluid,
\begin{equation}
\rho_i = \begin{cases}
\dfrac{4}{3(\omega_i+1)^2 \kappa T^2} & \textrm{if}\; \omega_i \neq -1\\
\Lambda/\kappa & \textrm{if} \; \omega_i = -1
\end{cases},
\end{equation}
where $T$ is the time coordinate in the FLRW metric. An extensive treatment of this computation can be found in the Appendix \ref{sec:Appendix_Cosmology}.

\subsection{Coordinate transformation and an application to the $\Lambda$CDM model}
\label{subsec:Coordinate transformation and an application to the LCDM model}

Let us consider the following situation: Two remote galaxies are merging so their central super-massive black holes are orbiting around a common center of mass and slowly approaching to each other. This is basically a Keplerian problem with --approximate-- spherical symmetry. For this reason, near the source of gravitational waves, the set of coordinates $\{t,r,\theta,\phi\}$ with spherical symmetry is very useful in order to describe spacetime and their perturbations. A very detailed discussion about these considerations are given in \cite{Espriu:2014mwa}, where also it is established that GWs in their simplest form and expressed in these source-centered coordinates will have the form of the equation \eqref{GW_solution_harmonic}.\\

However, these coordinates are not useful in cosmology because the cosmological measurements are described in comoving coordinates, i.e. $\{ T,R,\theta,\phi \}$. Thus, the main objective of this research is to find the coordinate transformation between $\{ t,r \}$ and $\{ T,R\} $ and thus, taking advantage of the principle of covariance and relativity, we will able to describe the propagation of gravitational waves as seen by a cosmological observer.  For a compendium with all the results found before the writing of this paper, including the coordinate transformations of a de Sitter Universe, besides other implications, see \cite{Alfaro:2017sxd}. \\

The easiest example that we can give to illustrate the situation is showing the de Sitter case, where only the action of the cosmological constant is taken into account. We note that in a vacuum background, an approximately spherical source of GWs would produce a Schwarzschild metric. Thus, if we take $\Lambda\neq 0$, then when we are far from the source (i.e. neglecting the mass term at the cosmological horizon, where $\Lambda r^3 \gg 6M$), the geometry of spacetime will be described approximately by a de Sitter (dS) metric,
\begin{equation}
\label{SdS_metric_mod}
\dd s^2 = - \left( 1  - \frac{\Lambda}{3} r^2 \right) \dd t^2 + \frac{\dd r^2}{1 - \dfrac{\Lambda}{3} r^2} + r^2 \dd \Omega^2.
\end{equation}

On the other hand, the FLRW metric --expressed in comoving coordinates-- will be given by \eqref{FLRW_metric} and the scale factor is of the form $a(T)=a_0\exp(\sqrt{\Lambda/3}\Delta T)$ (see appendix \ref{sec:Appendix_Cosmology}). The main idea is to express the coordinates of the SdS metric in terms of the comoving coordinates of the FLRW metric, as they are two equivalent representations of the same spacetime. It was found that the coordinate transformation is given by,
\begin{align}
r(T,R) &= a(T) R \label{SdS:tran_a}\\
t(T,R) &= T - \sqrt{\frac{\Lambda}{3}} \ln \sqrt{ 1 - \frac{\Lambda}{3} a(T)^2 R^2 }. \label{SdS:tran_b}
\end{align}	

If we expand them at order $\sqrt{\Lambda}$, we get, 
\begin{align}
r(T,R) &= a_0 R\left[ 1 + \Delta T \sqrt{\frac{\Lambda}{3}}\right] + \order{\Lambda} \label{SdS:trans_lin_b}\\
t(T,R) &= T + a_0^2 \left( \frac{R^2}{2} \sqrt{\frac{\Lambda}{3}}\right) + \order{\Lambda}.\label{SdS:trans_lin_a}
\end{align}

This result was obtained previously \cite{Bernabeu:2011if,Espriu:2012be}, and it was the starting point of this line of work. From these transformations it was found how the cosmological constant affects on  the propagation of GWs and it was explained how this effect could be measured using PTAs \cite{Espriu:2014mwa}. These expressions show us how comoving coordinates are related to the coordinates in the dS metric, and replacing the coordinates in \eqref{GW_solution_harmonic} would show how GWs are seen by a cosmological observer. That is the main idea and it is what we are going to exploit next.\\

In order to develop a more general discussion of the phenomenon, we will first consider a Universe filled by a single fluid with an arbitrary equation of state, i.e. $p_i = \omega_i \rho_i$. The methodology to be used is basically build a diagonal, spherically symmetric and asymptotically flat metric (that we will denote by SS$\omega_i$), described in the coordinates $\{t,r\}$, that recovers the corresponding FLRW metric in comoving coordinates, then find the coordinate transformation between both frames and, finally, replace the coordinates in \eqref{GW_solution_harmonic}, showing how it affects on the propagation of gravitational radiation. \\

In the appendix \ref{sec:Appendix_SSw_metric} is available the full derivation of the exact expression of the the SS$\omega_i$ metric, which is of the form
\begin{equation}
\dd s^2 = -\dfrac{\dd t^2}{\left(1-\dfrac{\kappa\rho_{i}r^2}{3} \right)\left( 1 + \dfrac{\kappa\rho_{i} r^2 (3\omega_i+1)}{6}  \right)^{\dfrac{1-3\omega_i}{1+3\omega_i}}} + \dfrac{\dd r^2}{1-\dfrac{\kappa\rho_{i}r^2}{3}} + r^2 \dd \Omega^2 ,
\end{equation}
and the coordinate transformations between $\{t,r\}$ and $\{T,R\}$ in terms of $\rho_{i}$ and $\rho_{0}=\rho_i(T_0)$, which are
\begin{subequations}
	\begin{align}
	t &= \frac{\left[ c + R^2 (\kappa \rho_{0})^{\frac{2}{3(\omega_i+1)}}(\kappa \rho_i)^{\frac{3\omega_i+1}{3(\omega_i+1)}}  \right]^{\frac{1}{2n}}}{\left(A^{\frac{1}{2n}} \right)\sqrt{\kappa \rho_{i}}} \label{SSw:Final_transformations_A} \\
	r &= a(T)R=R\left( \frac{\rho_{0}}{\rho_{i}}  \right)^{\frac{1}{3(\omega_i+1)}}, \label{SSw:Final_transformations_B}
	\end{align}
\end{subequations}

where $A$ and $n$ are constants (see appendix \ref{sec:Appendix_SSw_metric}). The expansion of \eqref{SSw:Final_transformations_A} and \eqref{SSw:Final_transformations_B} in terms of the energy density of the fluid at the present day, i.e. $\rho_{0}$, becomes
\begin{subequations}
\begin{align}
t &= T + \frac{R^2}{2} \sqrt{\frac{\kappa \rho_{0}}{3}}  + \frac{R^2}{12}(1-3\omega_i) \kappa\rho_{0} \Delta T+ \order{\kappa^2 \rho_{0}^2} \label{SSw:Linear_transformations_A} \\
r &= R \left( 1 + \Delta T \sqrt{\frac{\kappa \rho_0}{3}} -\frac{\kappa \rho_{0}\Delta T^2}{12} (1+3\omega_i)  \right) + \order{\kappa^2\rho_{0}^2}. \label{SSw:Linear_transformations_B} 
\end{align}	
\end{subequations}

These results agree with the previous works \cite{Espriu:2012be,Espriu:2014mwa,Alfaro:2017sxd}, and they show that, regardless the equation of state, the first term in the expansion is always at order $\sqrt{\rho_0}$. Using this fact, we can expand the First Friedmann equation at first order in $H_0$,
\begin{equation}
\label{linear_scale_factor}
a(T) = 1 + H_0 \Delta T + \order{H_0^2},
\end{equation}
where $H_0$ is the \textit{Hubble constant}, given by
\begin{equation}
H_0 = \sqrt{ \frac{\kappa \rho_{\textrm{eff}}(T_0)}{3}} = \sqrt{\frac{\Lambda}{3} + \frac{\kappa \rho_{d0} }{3} + \frac{\kappa \rho_{r0}}{3}}.
\end{equation}

As the $r$ coordinate has to transform as $r \to a(T) R$ to preserve spherical symmetry, a comparison between \eqref{SSw:Linear_transformations_A}, \eqref{SSw:Linear_transformations_B} and \eqref{linear_scale_factor} shows that the cosmological components are added inside the square root, as was discussed by \citet{Alfaro:2017sxd} in the case of non-relativistic matter. Thus, in order to obtain the correct limits for the previous models, the most general linearized coordinate transformations must be of the form,
\begin{subequations}
	\begin{align}
	t &= T + \frac{R^2}{2} H_0+ \order{H_0^2}\label{LCDM_linear_transformations_A} \\ 
	r &= R\left(1 + \Delta T H_0\right) + \order{H_0^2}. \label{LCDM_linear_transformations_B}
	\end{align}
\end{subequations}

Note that these are linear in $H_0$, so only small effects  will be considered. By replacing \eqref{LCDM_linear_transformations_A} and \eqref{LCDM_linear_transformations_B} into \eqref{GW_solution_harmonic}, we obtain an expression in terms of the comoving coordinates,

\begin{equation}
\label{GW_solution_anharmonic}
{h'}_{\mu \nu}^{\textrm{(GW)}} =  \frac{\left( 1 + R H_0 \right)}{R}  A'_{\mu \nu}\sin \left[-w_{\textrm{eff}} T + k_{\textrm{eff}} R \right] + \order{H_0^2},
\end{equation}
where $A_{\mu\nu}'$ are the transformed components of the polarization tensor, and the effective angular frequency and wave number are given by
\begin{equation}
\label{redshift}
w_{\textrm{eff}}\, \equiv \,\Omega \left( 1 - R H_0\right),\quad k_{\textrm{eff}} \,\equiv \,\Omega \left(1 - \frac{R}{2} H_0 \right).
\end{equation}

From the last two expressions we can infer how the Hubble constant affects the propagation of GWs when a cosmological observer (e.g. laboratories in the surface of the Earth or local celestial bodies as pulsars) is measuring them, using comoving coordinates. These results show that the previous findings \cite{Bernabeu:2011if,Espriu:2012be,Espriu:2014mwa,Alfaro:2017sxd} were merely approximated and, therefore, incomplete. At the same time, we discard the idea of the possibility of measuring the cosmological constant $\Lambda$ separately from the other components of the Universe, since all of them are coupled within $H_0$. It is important to note that the expression for the effective frequency $w_{\textrm{eff}}$ in the equation \eqref{redshift} reproduces the usual cosmological redshift expected from the expansion of the Universe. However, the effect on the wave number $k_{\textrm{eff}}$ cannot be derived from other simpler considerations, e.g. time dilation or redshift, and represents an additional feature of this framework. On the other hand, as it is discussed by \citet{Alfaro:2017sxd}, even when the phase velocity of the GW is not exactly $1$ (in natural units), if it is computed with respect to the ruler distance traveled, it can be shown that its value is exactly equal to $1$.

\section{Pulsar Timing Arrays}
\label{sec:Using PTA to measure the Hubble constant}
\subsection{Timing residual and the working of PTAs}
\label{subsec:Timing residual and the working of PTA}

The results obtained in the last section, e.g. equations \eqref{GW_solution_anharmonic} and \eqref{redshift}, show that the Hubble constant should influence on the propagation of Gravitational Waves. Now we will set an experimental framework in which this effect can be eventually measured. For this, we will use the light coming from a local pulsar and the shift in the time of arrival of the electromagnetic (EM) pulse due to the pass of GWs. In the following picture we show the simplest configuration, which will guide our discussion.\\

\begin{figure}[h!]
	\centering
	\includegraphics[width=0.5\textwidth]{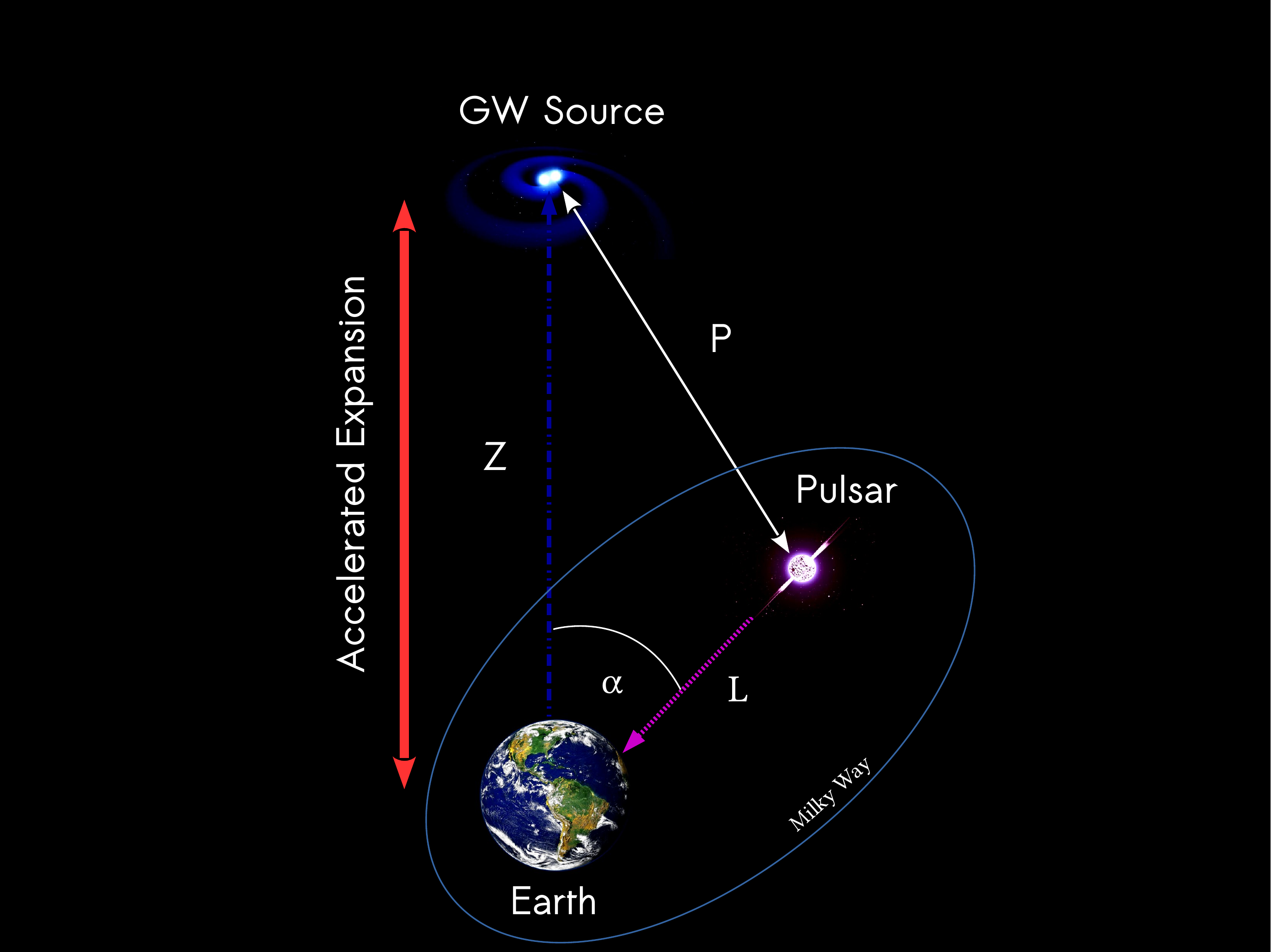}
	\caption{Setup of the configuration of our study: A source of gravitational waves ($R=0$), the Earth (at distance $Z$) and a nearby Pulsar located at $\vb{P} = (P_X,P_Y,P_Z) $ referred to the source. The Z direction is chosen to be defined by the source-Earth axis. Polar and azimuthal angles are $\alpha$ and $\beta$ respectively, from Z axis (self-elaborated image).\label{fig:setupfinal}}
\end{figure}

From figure \ref{fig:setupfinal}, we note that the Earth and the pulsar are gravitationally bounded to the Milky Way, so they do not feel the Universe's expansion. However, the source of GWs and the system Earth-pulsar are not bounded, so they do feel accelerated expansion and, therefore, the discussion and results of the previous chapter apply. The pulsar emits light with a particular EM field. Denoting the time-dependent phase of this field at the pulsar as $\phi_0$, then the phase of the EM pulse measured from Earth can be expressed as
\begin{equation}
\label{EM field phase}
\phi(T) = \phi_0\left[T - \frac{L}{c} - \tau_0(T) - \tau_{\textrm{GW}}(T)\right]
\end{equation}
where $c$ the speed of light, $\tau_0(T)$ is the timing correction associated to the motion of the Earth respect to the Solar system and $\tau_{\textrm{GW}}(T)$ is the timing correction due to the action of GWs passing through the system.   The correction due to the action of Gravitational Waves is given \citep{Finn2009,Deng2011,Espriu:2014mwa,Alfaro:2017sxd} by the following expression
\begin{equation}
\label{tau_GW}
\tau_{\textrm{GW}} (T) = -\frac{1}{2} \hat{n}^i \hat{n}^j H_{ij}(T),
\end{equation}  
where $\hat{n} = (-\sin \alpha \cos \beta, -\sin\alpha\cos\beta,\cos\alpha)$ is a unit vector pointing from Earth to the pulsar and $H_{ij}$ is the integral of the metric perturbation along the null geodesic in the path pulsar--Earth, which could be parameterized by $\vec{R}(x)=\vec{P} + L(1+x)\hat{n}$ with $x\in[-1,0]$.  Using this path, $H_{ij}(T)$ takes the following form

\begin{equation}
\label{Hij}
H_{ij}(T) = \frac{L}{c} \int_{-1}^{0} {h'}_{\mu \nu}^{\textrm{(GW)}} \left(T + \frac{L}{c} x , \abs{\vec{R}(x)} \right) \dd x.
\end{equation}

\subsection{Including the $\Lambda$CDM model}
\label{subsec:Including the LCDM model and numerical analysis}

In our framework (see figure \ref{fig:setupfinal}), the source of GWs is far away from Earth, though the pulsar is at a local distance. Therefore, we can consider the reasonable approximation $L/Z \ll 1$ and using it we can show that $R(x)\approx Z + xL\cos\alpha$. Now we compute the equation \eqref{tau_GW} using \eqref{Hij}. However, in the TT-Lorenz gauge, for a GW propagating through the Z axis, the only non-zero values of the components of $A'_{\mu \nu}$ are in the X,Y components \citep{Bernabeu:2011if}. We can also additionally assume, in order to simplify the computation of the integral, that $\abs{A'_{\mu \nu}} \equiv \varepsilon\; \forall \mu,\nu$. Thus, the full timing residual in the arrival time of the pulsar due to the pass of GWs in the $\Lambda$CDM model becomes,

\begin{align}
\label{tau_GW_LCDM_full}
\tau_{\textrm{GW}}^{\Lambda\textrm{CDM}}  = - \frac{L\varepsilon}{2c} (\sin^2\alpha\cos^2\beta + &2\sin^2\alpha\cos\beta\sin\beta-\sin^2\alpha\sin^2\beta) \\
&\times \int_{-1}^{0} \frac{1 + H_0 \left[ T_e + \frac{xL}{c} \right]   }{Z + xL\cos\alpha} [\cos\Theta + \sin\Theta] \; \dd x , \nonumber
\end{align}
where
\begin{align}
\label{Theta_LCDM_full}
\Theta = \Omega \left\{ \left( 1 - \frac{Z+xL\cos\alpha}{c} H_0 \right) \left(T + \frac{xL}{c}\right) - \left( 1 - \frac{Z+xL\cos\alpha}{2c} H_0 \right) \left(\frac{Z+xL\cos\alpha}{c}\right) \right\}.
\end{align}

\section{Results and discussion}
\label{sec:results_discussion}
\subsection{Simulation of the timing residual of an individual pulsar}
\label{subsec:simulation}
From a geometric argument, we can always fix the pulsar, the Earth and the source of gravitational waves in the same plane, thus we can set $\beta = 0$, and the geometrical parameters involved in $\tau_{\textrm{GW}}$ are the angle $\alpha$ between the Pulsar and the GW-Source, the distance Earth--Pulsar, $L$, and the distance Earth--GW-source, $Z$. In order to perform a numerical analysis we can choose some reasonable values of the parameters that appear in the equation \eqref{tau_GW_LCDM_full} and fix them to visualize the behavior of the timing residual $\tau_{\textrm{GW}}$. Thus, the setup described in the figure \ref{fig:setupfinal} can be approximately modeled with the values that appear in the Table \ref{tab:parameters}.

\begin{table}[h!]
	\centering
	\begin{tabular}{|c|cc|}
		\hline
		Parameter&SI value& \\
		\hline
		$Z$ & $3 \times 10^{24}$ m & $\sim100$ Mpc \\ 
		$T$ & $Z/c = 10^{16}$s & $\sim300$ Myr \\ 
		$L$ & $10^{19}$m & $\sim1000$ ly \\ 
		$\Omega$ & $10^{-8}\, \textrm{rad}/s$ &  \\ 
		$\varepsilon$ & $1.2\times 10^{9}$ m &  \\ 
		\hline
	\end{tabular}
	\caption{\label{tab:parameters} List of values considered for the parameters in the numerical integration of the timing residual $\tau_{\textrm{GW}}$ in \eqref{tau_GW_LCDM_full}, according to current accuracy of PTAs.}
\end{table}

\begin{figure}[h!]
	\centering
	\subfloat[\label{fig:timing_dif}]{%
		\includegraphics[width=0.45\linewidth]{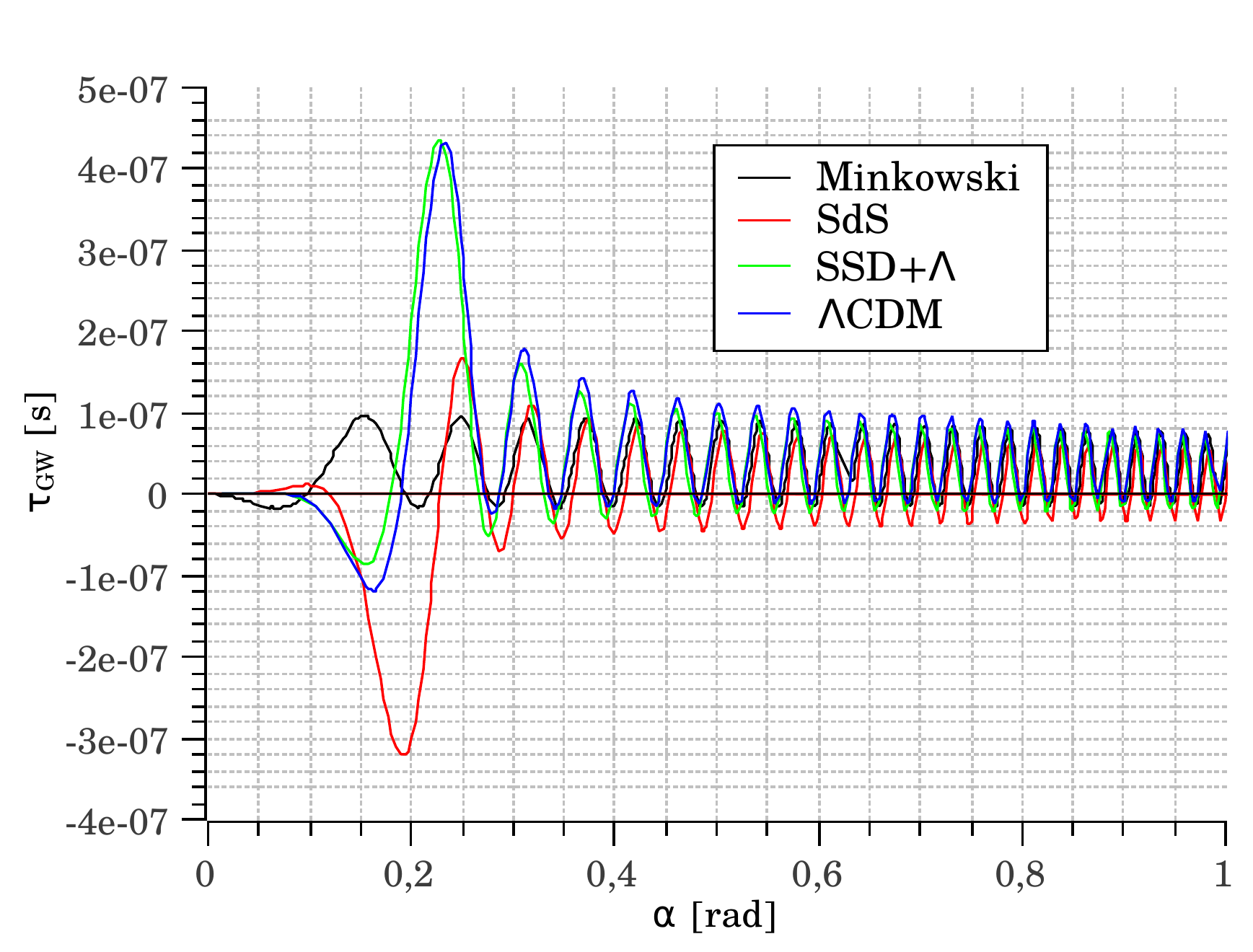}%
	}
	~ 
	\subfloat[\label{fig:H0values}]{%
		\includegraphics[width=0.45\linewidth]{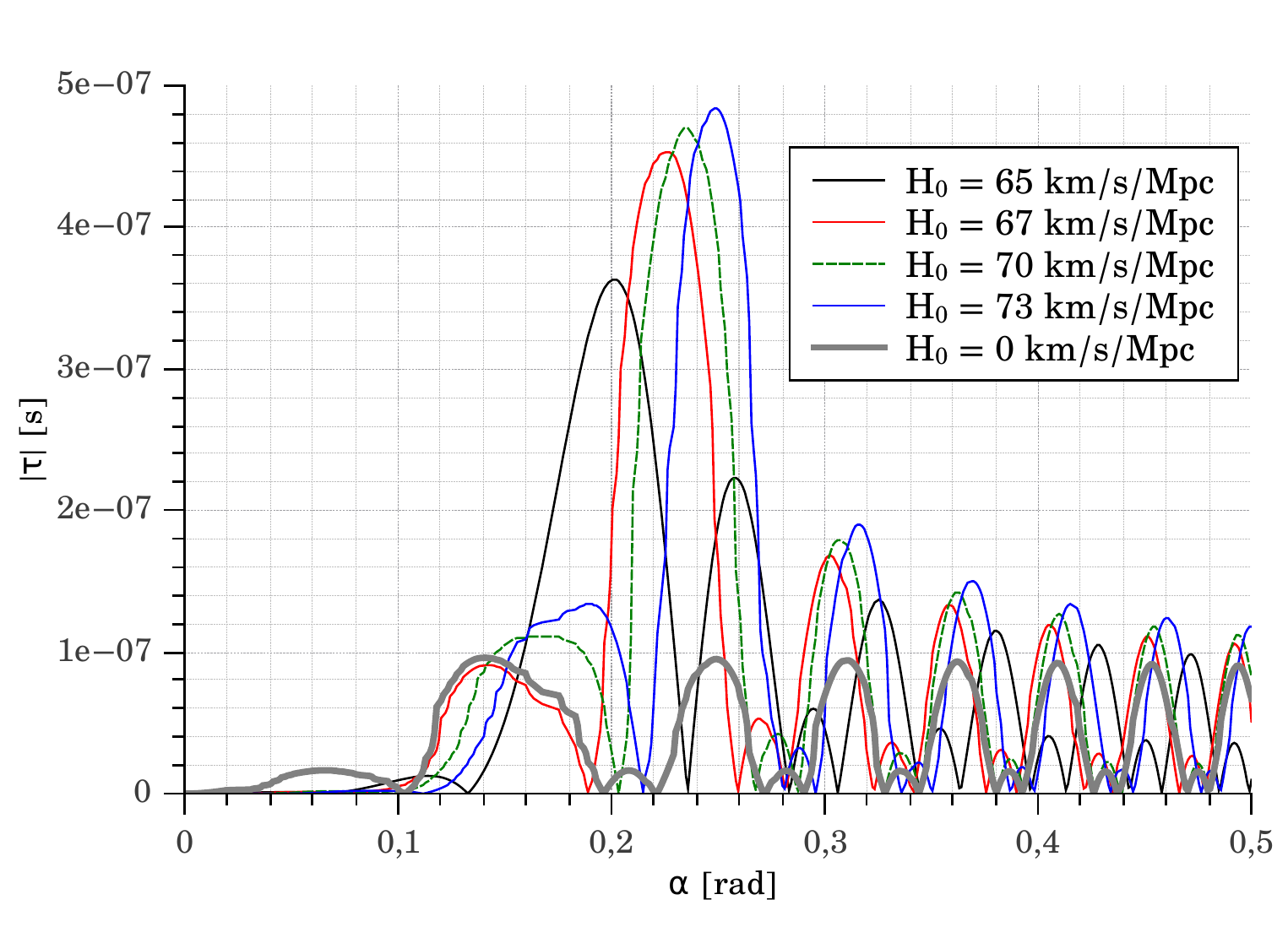}%
	}
	\caption{(a) Comparison between different material contents of the Universe. SdS is the de Sitter case, SDS$+\Lambda$ is where Dark Energy and Dark Matter (dust+$\Lambda$) are taken account. The $\Lambda$CDM case also includes radiation. Note that in the Minkowski spacetime no peak is observed. This graphic also agrees with the results obtained by \citet{Alfaro:2017sxd}. (b) Numerical analysis of the absolute value of timing residual in terms of $\alpha$, varying the value of $H_0$. For a non--zero $H_0$, a dominant peak is present, whose angular position (i.e. at an angle $\alpha_m$) in the $\alpha$--axis increases as the value of $H_0$ also increases.}
\end{figure}

For the source of GWs, we choose a typical distance $Z$ where supermassive black holes are present and although is large, it is not a cosmological distance near to the Big Bang. On the other hand, the distance between Earth and the pulsar is within the margin of a local galactic scale. It can be seen, from table \ref{tab:parameters}, that $L\ll Z$, as required from the previous considerations. The angular frequency is of the expected order for future PTA projects and the same argument is used to fix $\varepsilon$, due to that it satisfies $\abs{h}\sim \varepsilon/R\sim 10^{-15}$, where $\abs{h}$ and $\Omega$ are within the expected accuracy of PTA projects, e.g. the EPTA \citep{Babak2016} or the NANOGrav collaboration \citep{Nano2019_mm}. Employing these parameters, the numerical integration of $\tau_{\textrm{GW}}$ gives the results that are shown in figure \ref{fig:timing_dif}. As we can see, the value of the timing residual can be positive or negative. Since the meaningful physical magnitude is the amount of time, rather than the direction of the shift, we can also plot the absolute value of $\tau_{\textrm{GW}}$, but now changing the value of $H_0$ within a region of parameters, obtaining the plot in figure \ref{fig:H0values}.

\subsection{Statistical significance and the timing residual analysis}
\label{subsec:statistical}
The most important feature of these figures is the presence of a considerable peak in the value of the timing residual for a certain angle $\alpha$. Moreover, we note that this peak changes its angular position with the value of the Hubble constant. This is our first clue of the existence of a distinguishable signal coming from the cosmological effects on the propagation of GWs.\\

In order to make an analysis of the possible signal shown in the previous figures, we will use some pulsars from the ATNF catalog \citep{Catalogo} and follow the ad hoc analysis of \citet{Espriu:2014mwa}. As we know, pulsars are stable clocks whose periods are known with great accuracy. Assuming a modest precision of $\sigma_t = 9.6 \times 10^{-7}s \approx 10^{-6}s$ which is obtained by averaging the precision achieved of best pulsars in the IPTA collaboration, we can define a \textit{statistical significance} of the timing residual, of the form
\begin{equation}
\sigma = \sqrt{\frac{1}{N_p N_t} \sum_{i,j=1}^{N_p,N_t} \left( \frac{\tau_{\textrm{GW}}}{\sigma_t} \right)^2},
\end{equation}   
where index $i$ running from 1 to $N_p$ (number of pulsars averaged) and $j$ running from 1 to $N_t$ (number of observations). Assuming we perform measurements every 11 days through 3 years, then $N_t = 101$. The pulsars belong to the considered cluster are shown in table \ref{tab:pulsars}.\\

\begin{table}[h]
	\centering
	\begin{tabular}{|c|cc|}
		\hline
		Pulsar Name&$\phi$&$L_i$\\
		\hline
		J0024-7204E & $-44.89^\circ$ & 4.69 kpc \\ 
		J0024-7204D & $-44.88^\circ$ & 4.69 kpc \\ 
		J0024-7204M & $-44.89^\circ$ & 4.69 kpc\\ 
		J0024-7204G & $-44.89^\circ$ & 4.69 kpc\\ 
		J0024-7204I & $-44.88^\circ$ & 4.69 kpc \\ 
		\hline
	\end{tabular}
	\caption{\label{tab:pulsars}List of pulsars averaged for an hypothetical source at angular separation $\alpha$. It is shown the data given in \cite{Catalogo}, where $\phi$ is the galactic latitude --transformed to $\beta_i$-- and $L_i$ the distance between Earth and pulsar. We can note that this set simplify the computation of $\sigma$ because all pulsars are very close to each other.}
\end{table}

We will keep $\alpha$ as a free parameter and suppose that an hypothetical GW source is located at $\alpha$ radians between Earth and pulsars. Thus, the statistical significance is given by
\begin{equation}
\sigma (\alpha) = \sqrt{\frac{1}{5 \cdot 101} \sum_{i=1}^{5} \sum_{j=1}^{101} \left( \frac{\tau_{\textrm{GW}}(\beta_i)}{\sigma_t} \right)^2 }
\end{equation}

\begin{figure}[h!]
	\centering
	\subfloat[\label{fig:statistical_raw}]{%
		\includegraphics[width=0.45\linewidth]{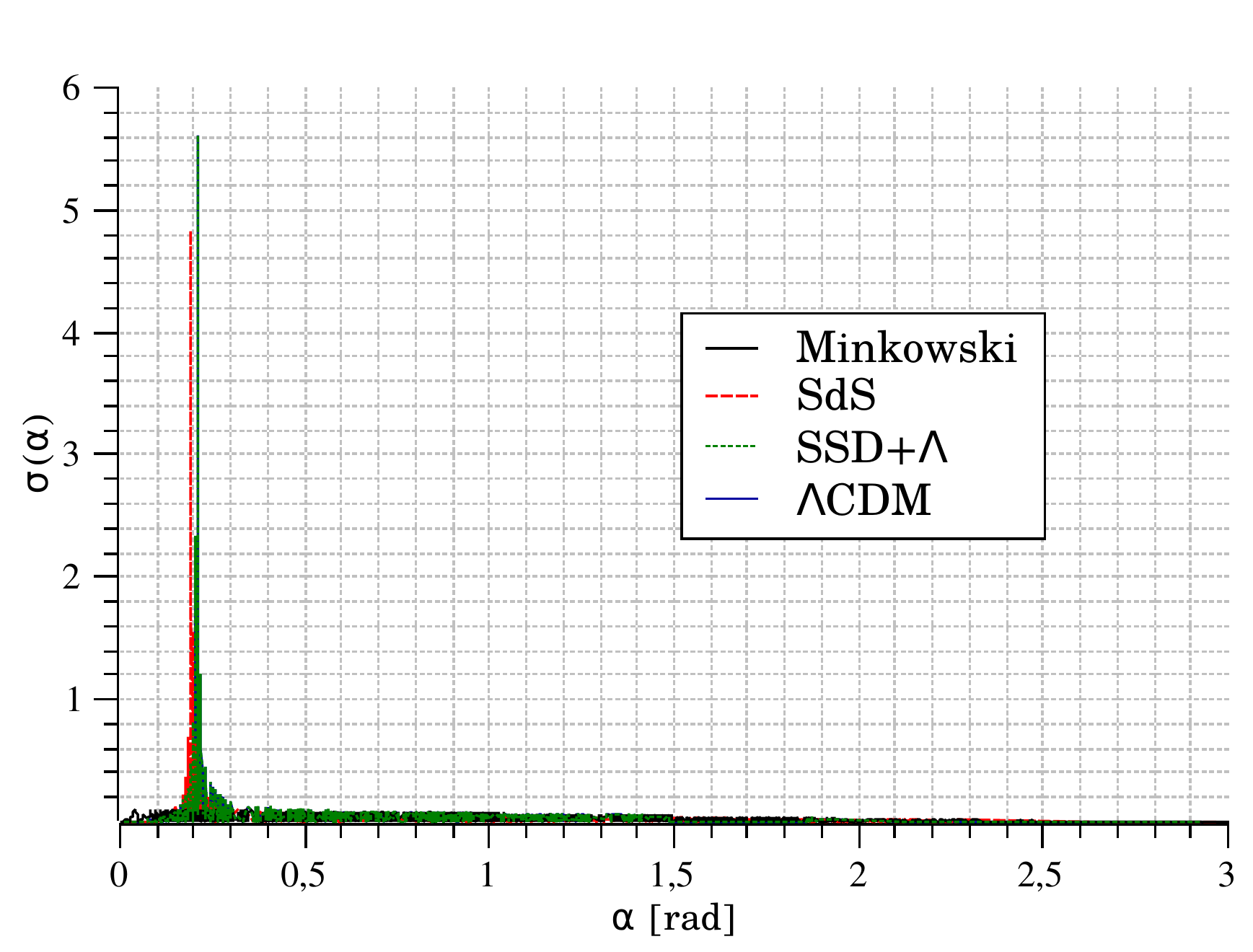}%
	}
	~ 
	\subfloat[\label{fig:statisticals}]{%
		\includegraphics[width=0.45\linewidth]{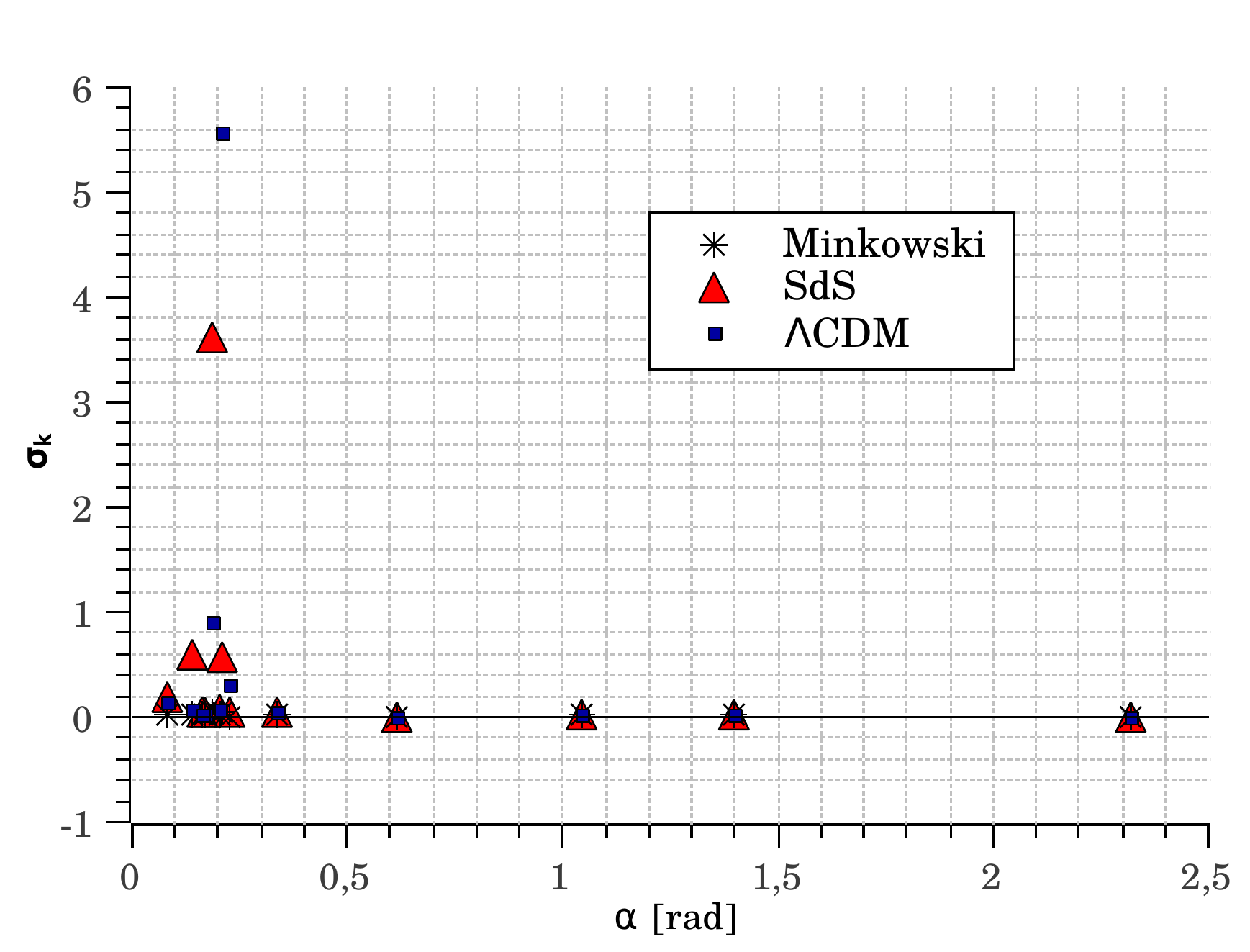}%
	}
	\caption{(a) Simplified simulation of $\sigma$ in an hypothetical observation of the peak in $\tau_{\textrm{GW}}$, which is located near 0.2 rad. Green and blue curves overlap due to the similarity of models. (b) Numerical simulation $\sigma$ in the measurement of $\tau_{\textrm{GW}}$ for three different models. We used 13 sets with 5 pulsars each, and for 11 of them, we took randomly distributed pulsars from the ATNF catalog (see appendix \ref{sec:Appendix_pulsars}), and 2 of them as test groups with suitable parameters. The larger peaks come from the later, showing the difficulty of a successful measurement.}
\end{figure}

The result of the simulation can be observed in the Figure \ref{fig:statistical_raw}, showing the characteristic peak as we expected. However, we can develop a more realistic simulation. In figure \ref{fig:statistical_raw}, only a cluster of 5 pulsars were considered and all of them were averaged at the same angle $\alpha$. However, one can expect that all the pulsars are located at different angles (in galactic coordinates) and basically being randomly located. Therefore, we have considered 11 randomly distributed groups of 5 pulsars each (see appendix \ref{sec:Appendix_pulsars}), two test clusters of pulsars with a suitable location (65 pulsars in total) and a source of gravitational waves located at galactic coordinates $\theta_S = 20^\circ$ and $\phi_S = 15^\circ$. \\

Then, we averaged them using the statistical significance given by
\begin{equation}
\sigma_k = \sqrt{\frac{1}{5\cdot 101} \sum_{i=1}^{5_k} \sum_{j=1}^{101} \left( \frac{\tau_{\textrm{GW}}(L_i,\alpha_i,\beta_i)}{10^{-6}}  \right)^2 }
\end{equation}
and plot it as a function of the average angle of the group, namely $\bar{\sigma}_k = \sum_{i=1}^{5_k} \alpha_i/5$. From this simulation we obtained figure \ref{fig:statisticals}, where we can note how the randomly distributed pulsars mostly do not show any signal, except for those that are located very close to the maximal angle, namely $\alpha_m$, where the value of the timing residual is maximized. For a pulsar located within the vicinity of the angle $\alpha_m$, the effect of the Hubble constant on the propagation of GWs and the capability of measure them using PTAs is greatly increased. \\

This fact indicates that only the pulsars placed near the angle $\alpha_m$ with respect to the source of GWs will show the characteristic peak in the timing residual with great statistical significance, which implies a major obstacle when trying to observe this effect. Nevertheless, as more pulsars are observed and studied, it is more likely to measure the existence of this peak, which could represent a challenge for the future astrophysical research of PTA experiments. \\

\begin{figure}[h!]
	\centering
		\subfloat[\label{fig:logOmega}]{%
		\includegraphics[width=0.42\linewidth]{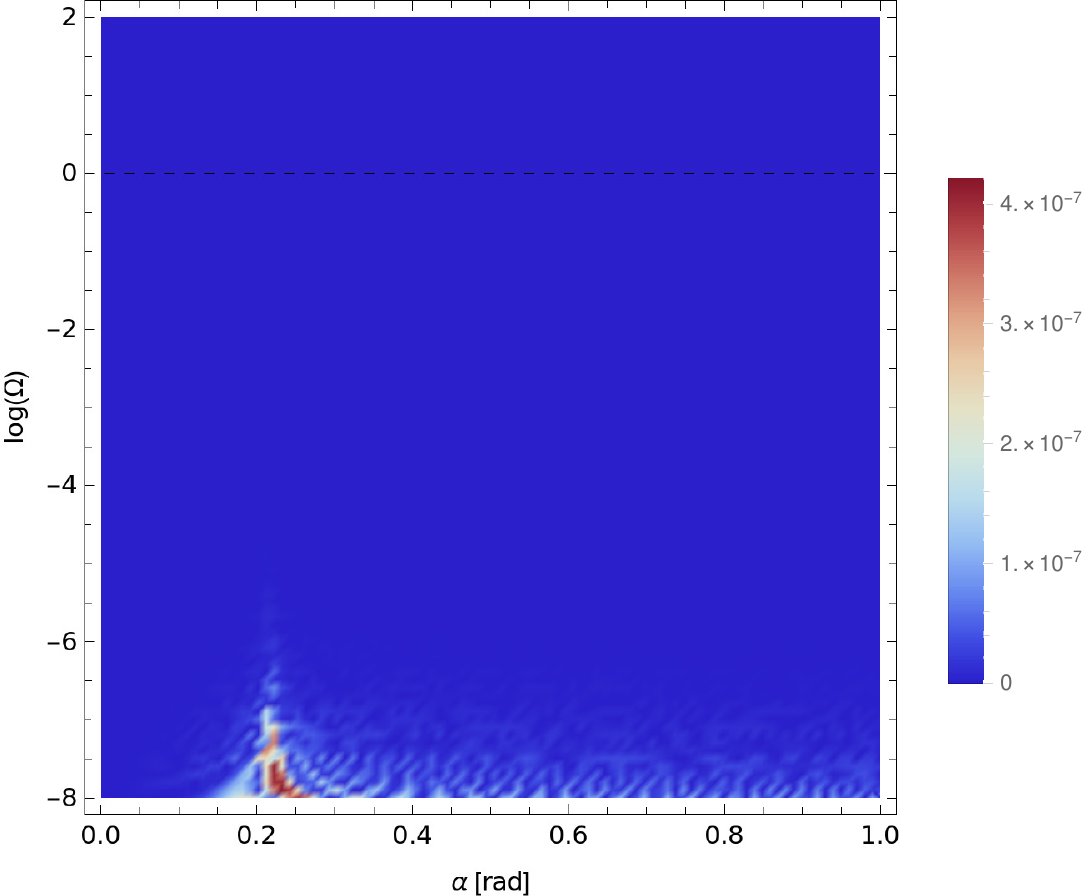}%
	}
	~ 
	\subfloat[\label{fig:Omega}]{%
			\includegraphics[width=0.45\linewidth]{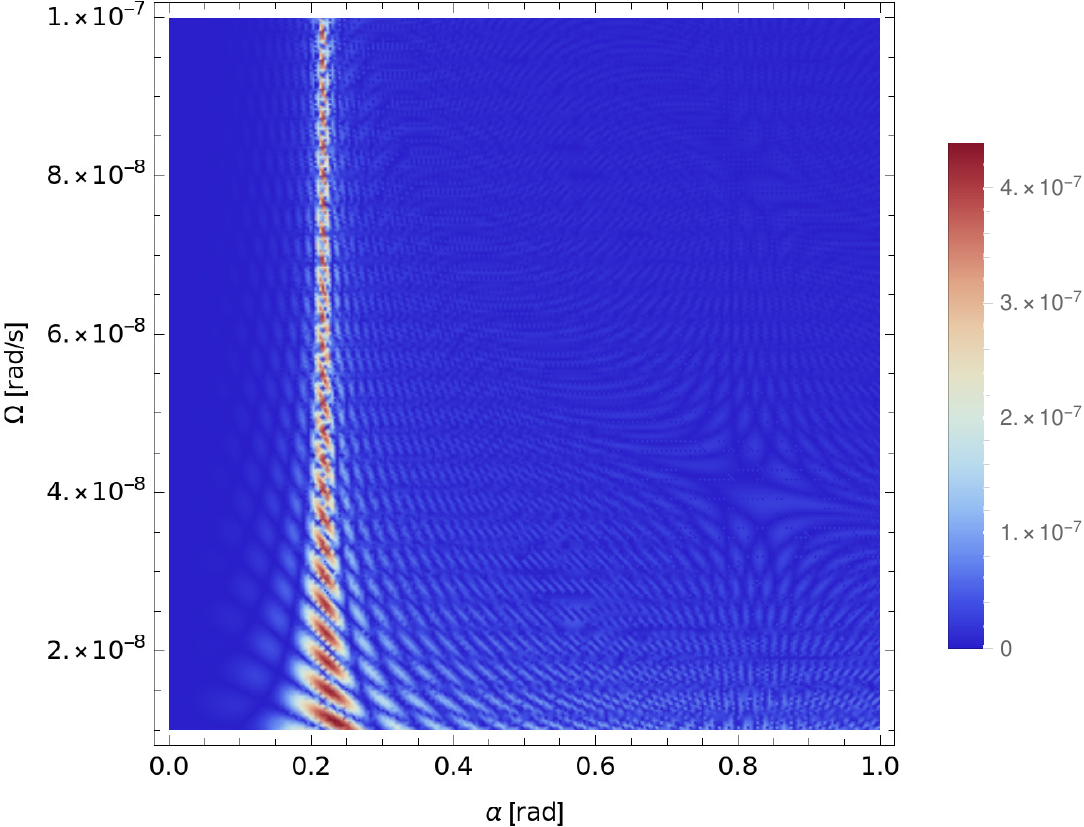}%
	}
	\caption{(a) Density plot of $\abs{\tau_{\textrm{GW}}}$ in terms of the common logarithm of angular frequency $\Omega$ and the angle $\alpha$. This graphic shows why PTAs are so important to measure this effect. Other values given by table \ref{tab:parameters}, with $H_0 = 70$ km/s/Mpc. (b) The same plot but focused in the range $10^{-8}\;\textrm{rad/s} < \Omega < 10^{-7}\;\textrm{rad/s}$. We can note the lack of dependence on $\Omega$.}
\end{figure}

First, in order to understand the role of the angle $\alpha_m$, in which the maximum of $\tau_{\textrm{GW}}$ is reached, we analyze the dependency on the original angular frequency of the incoming gravitational wave, namely $\Omega$, e.g. see eq. \eqref{Theta_LCDM_full}. From figure \ref{fig:logOmega} we note that for the region $10^{-6}\;\textrm{rad/s} < \Omega < 10^2\;\textrm{rad/s}$, the value of $\abs{\tau_{\textrm{GW}}}$ is practically zero. However, in the region $10^{-8}\;\textrm{rad/s} < \Omega < 10^{-6}\;\textrm{rad/s}$ the value starts to rise. This is the main reason why the other type of detectors as LISA or LIGO are useless in this context: Only PTA works in the proper range of the frequency spectrum \citep{NanoGrav2013,Barke2015,IPTA2016}. \\

\begin{figure}[ht!]
	\centering
			\subfloat[\label{fig:L_density}]{%
	\includegraphics[width=0.43\linewidth]{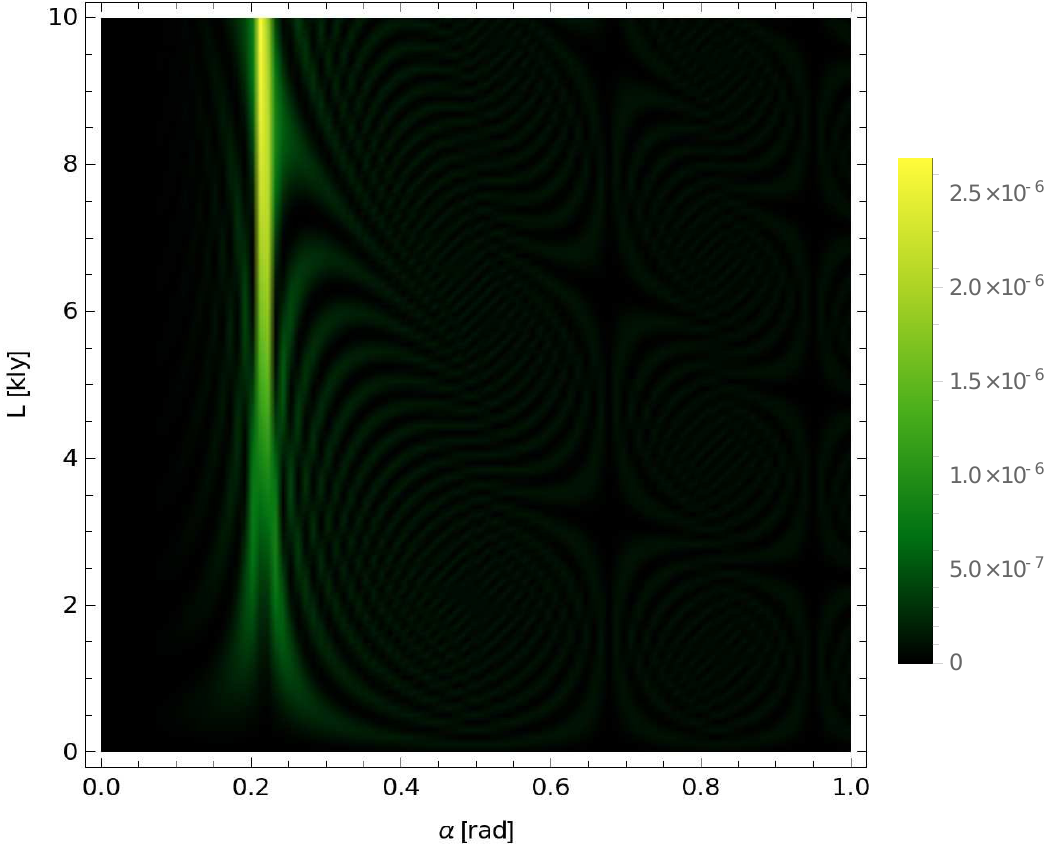}%
	}
	~ 
	\subfloat[\label{fig:Z_density}]{%
		\includegraphics[width=0.45\linewidth]{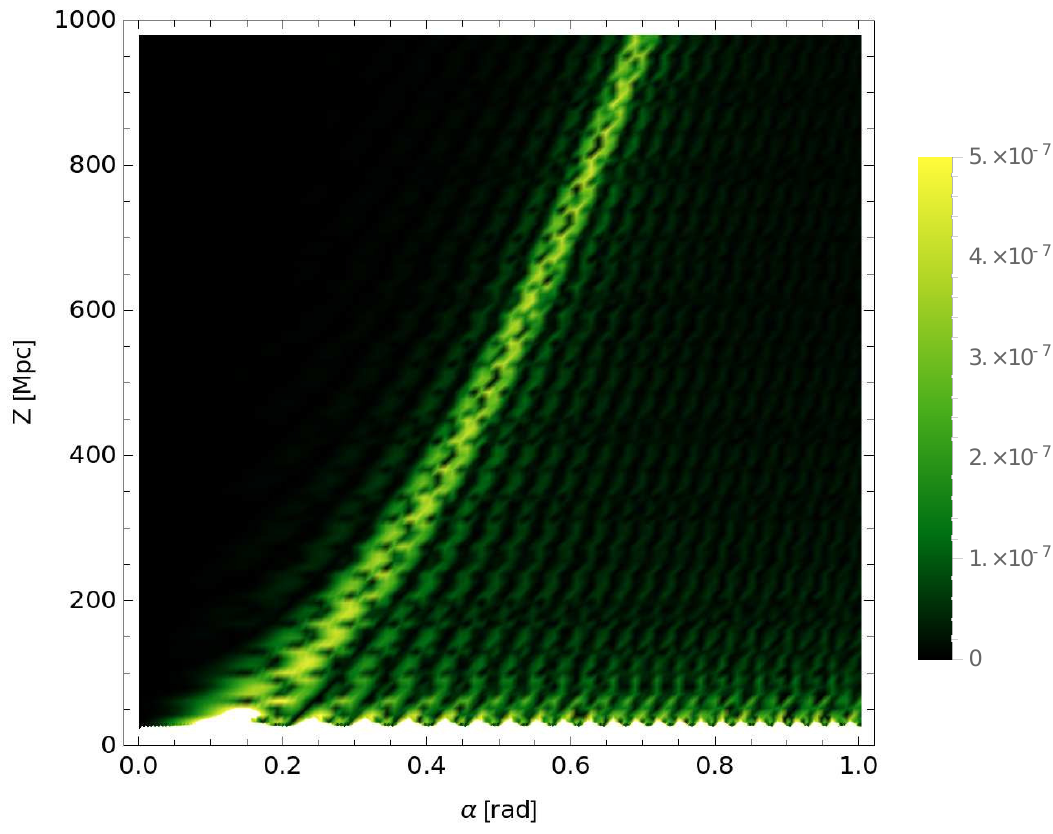}%
	}
	\caption{(a) Density plot of $\abs{\tau_{\textrm{GW}}}$ in terms of the distance $L$ (in kilolight-years), and the angle $\alpha$. The rest of parameters are given by table \ref{tab:parameters}, with $H_0 = 70$ km/s/Mpc. Again, there is almost no dependence on $L$. (b) Density plot of $\abs{\tau_{\textrm{GW}}}$ in terms of the distance $Z$ (in megaparsecs), and the angle $\alpha$. The rest of parameters are given by table \ref{tab:parameters}, with $H_0 = 70$ km/s/Mpc. Unlike the previous cases, we do see an angular dependency on $Z$.}
\end{figure}

In figure \ref{fig:Omega} we note a lack of angular dependence on the maximum values of $\tau_{\textrm{GW}}$, and moreover, the same behavior is observed for the distance $L$, in figure \ref{fig:L_density}.  \\

\begin{figure}[h!]
	\centering
				\subfloat[\label{fig:H0_500_density}]{%
		\includegraphics[width=0.42\linewidth]{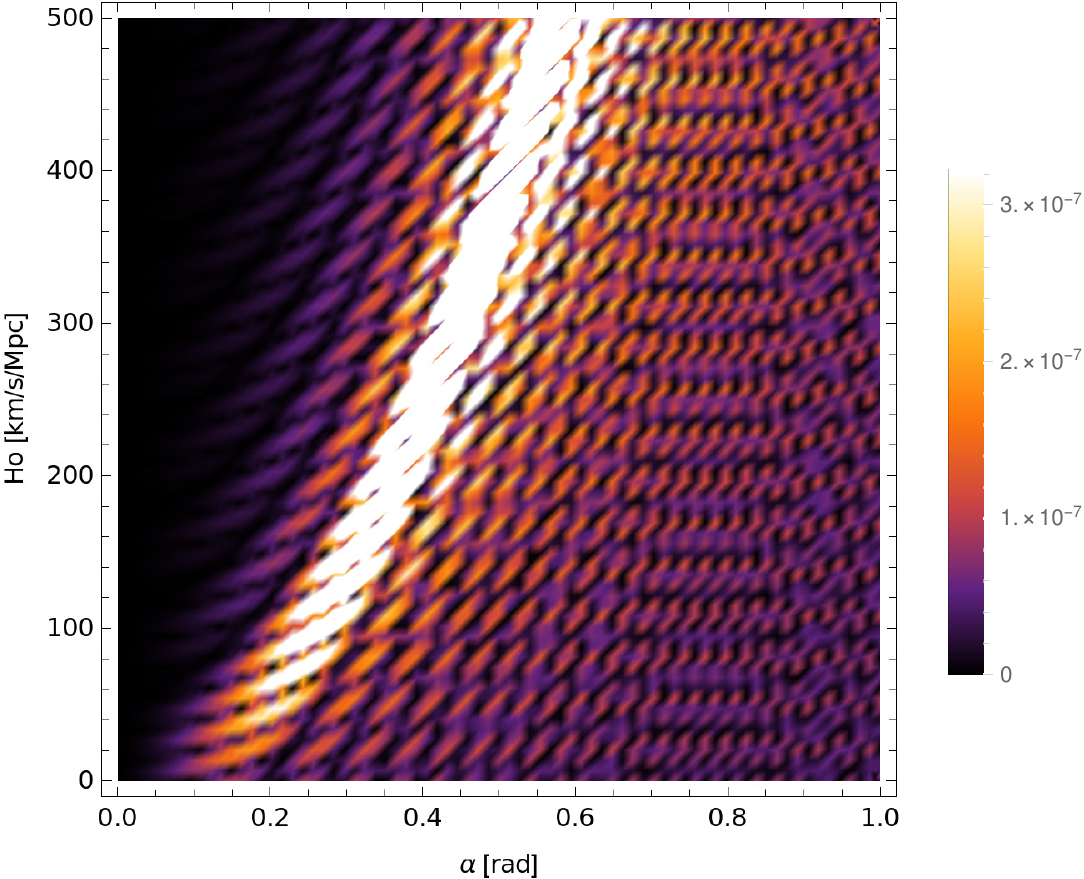}%
	}
	~ 
	\subfloat[\label{fig:H0_6080_density}]{%
		\includegraphics[width=0.42\linewidth]{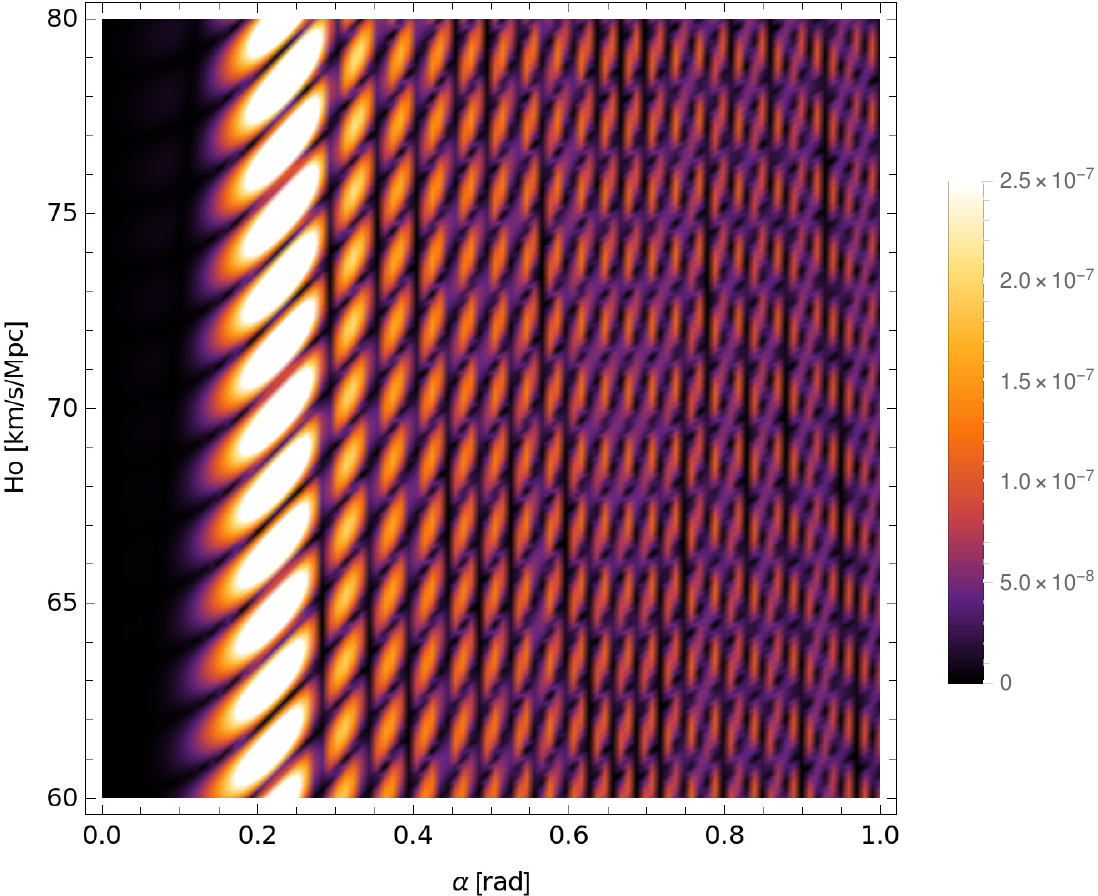}%
	}
	
	\caption{(a) Density plot of $\abs{\tau_{\textrm{GW}}}$ in terms of $H_0$, and the angle $\alpha$. The rest of parameters are given by table \ref{tab:parameters}. (b) The same plot but zoomed to the suitable range $60\;\textrm{km/s/Mpc}< H_0 <80\;\textrm{km/s/Mpc}$. We note a slight slope in angular position, in accordance with figure \ref{fig:H0values}.}
\end{figure}

However, in figure \ref{fig:Z_density} it is clear that for different values of the distance Earth--GW-Source, the angle $\alpha_m$ changes dramatically and therefore it depends on the distance $Z$ in an explicit but unknown way. For the Hubble constant, figure \ref{fig:H0_500_density} shows a similar situation. In fact, it represents how the dependency on $H_0$ is actually quite similar to the case of $Z$ in figure \ref{fig:Z_density}. From this behavior we can speculate about a possible relationship between these parameters. Zooming up, we obtain the figure \ref{fig:H0_6080_density}, where we can see that the maximum value of the timing residual $\tau_{\textrm{GW}}$ (i.e. the white spots) has a slightly oscillatory structure around a characteristic maximal angle $\alpha_m$. It can be noticed that the full white strip has a very small slope, showing a slow variation of the angular position, in agreement with the figure \ref{fig:H0values}.  \\

We can summarize the previous analysis as follows: The numerical integration of the equation \eqref{tau_GW_LCDM_full}, with the parameters given by the Table \ref{tab:parameters}, gives us the density plots shown above. From figure \ref{fig:logOmega} we can establish the crucial role of PTAs in the eventual measurement of the effects caused by the cosmological components of the Universe on the propagation of gravitational waves: The kind of effect that Dark Matter, Dark Energy and others cosmological fluids induce to the propagation of GWs is sensitive to the frequency spectrum of current Pulsar Timing Arrays experiments. \\

Furthermore, the figures \ref{fig:Z_density} and \ref{fig:H0_500_density} support the hypothesis of the existence of an implicit relationship between the $H_0$, $Z$, and $\alpha_m$. From the figures \ref{fig:timing_dif} and \ref{fig:H0values} we note that it could be difficult to find the best value of $H_0$ that fit the data unequivocally, due to the complicated dependencies of the parameters within the equation \eqref{tau_GW_LCDM_full} and because of the background (i.e. the null test for $H_0$, the flat Minkowski spacetime) has a not negligible amplitude for a large part of the angular values of $\alpha$. However, this situation does not happen within the vicinity of the characteristic peak of $\tau_{\textrm{GW}}$ when $H_0\neq 0$, which occurs at a certain angle $\alpha_m$, where the difference with the background is considerable. In that sense, if this tiny effect is ever observed, it is very likely that must be the case of a pulsar located in the vicinity of the angle $\alpha_m$, in such a way that the value of its timing residual is maximized, and that the signal is strong enough to rule out the experimental noise. In the section \ref{subsec:a_test_of_LCDM} the relationship between $H_0$, $\alpha_m$ and $Z$ will be clarified, and we could state the importance of it.

\subsection{A relationship between PTA observables and the Hubble constant}
\label{subsec:A relationship between PTA observables and the Hubble constant}

Using the stationary phase approximation for $\tau_{\textrm{GW}}$ and considering reasonable asymptotic expansions, we obtain the following expression,
\begin{equation}
\label{H0_approximation}
H_0 \cong \frac{2c}{Z} \sin[2](\frac{\alpha_m}{2}) .
\end{equation}

\begin{figure}[h!]
	\centering
	\includegraphics[width=0.4\linewidth]{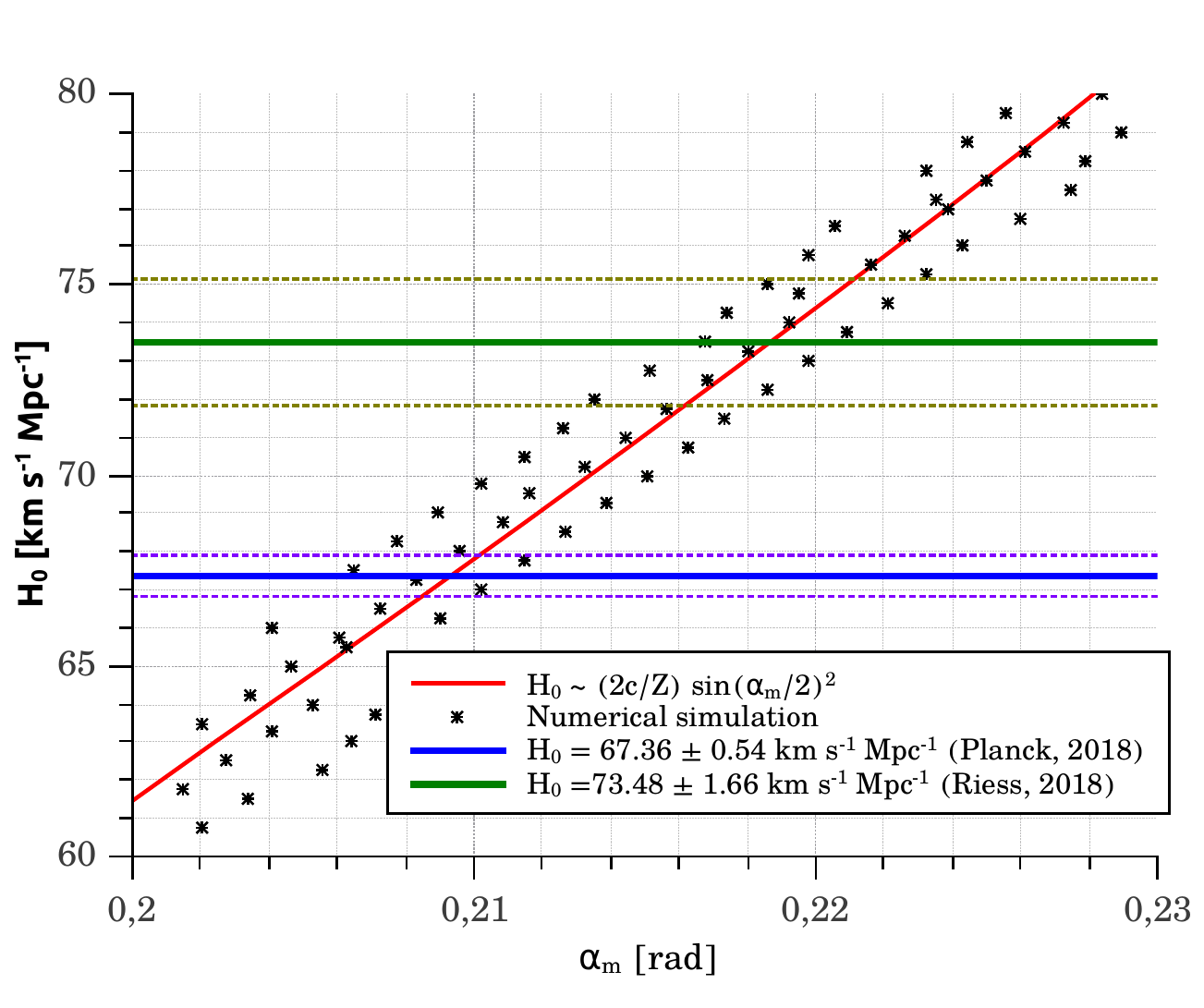}
	\caption{The value of $H_0$ using the formula \eqref{H0_approximation} and the numerical maximum of $\abs{\tau_{\textrm{GW}}}$. The average error in the approximation is of the 1.5\% from numerical simulation.}
	\label{fig:aproxcomp}
\end{figure}

A derivation of this equation can be found in appendix \ref{sec:Appendix}. The expression in \eqref{H0_approximation} provides a precise relationship between $H_0$ and the observables $\alpha_m$ and $Z$. Under ideal assumptions, this formula can be used to estimate the local value of the Hubble constant knowing the two main observables: $Z$ and $\alpha_m$. In figure \ref{fig:aproxcomp} we show the behavior of the approximation formula with respect to the numerical analysis, included some values of current observations.

However, there are some experimental obstacles when we want to measure $H_0$ using this framework, in particular because we need to determine the value of $Z$ by independent astrophysical methods. Nevertheless, a PTA experiment could determine, considering the experimental uncertainties, the value of the timing residual $\abs{\tau_{\textrm{GW}}}$, the value of the frequency of the source $\omega_{\textrm{eff}}$ and the direction of the incoming Gravitational Wave (and therefore, the value of $\alpha$). In the ongoing PTA experiments,  several pulsars are observed at once. Each of them will be located at different angular positions. If a GW is passing through the set of pulsars, then each of them will have, according to our model, a different associated $\abs{\tau_{\textrm{GW}}}$. Evidently, it is more likely that the pulsars have a relative angular position with the source in which the $\abs{\tau_{\textrm{GW}}}$ is not maximum (i.e. $\alpha \neq \alpha_m$). Clearly, these cases are very difficult to measure experimentally, because the value of $\abs{\tau_{\textrm{GW}}}$ for these cases is indistinguishable from the background. However, there is a tiny possibility that one or some of these pulsars are located just at an angular position where the value of $\abs{\tau_{\textrm{GW}}}$ is clearly distinguishable with respect to the background (i.e. $\alpha \approx \alpha_m$). If that were the case, then the value of $\abs{\tau_{\textrm{GW}}}$ should be large enough to be measured by the expected accuracy for PTA experiments.

\subsection{A test of the $\Lambda$CDM model using PTAs}
\label{subsec:a_test_of_LCDM}
With the spirit of testing the standard cosmological model, we know that for a small redshift, i.e. $z\ll 1$, it holds that $z \approx (Z/c) H_0$ \cite{Ryden}. Thus, when we compare this known formula with \eqref{H0_approximation}, we obtain an expression that could be interpreted as the redshift of the source of GWs in terms of the maximal angle $\alpha_m$,
\begin{equation}
\label{rs}
z \cong 2 \sin[2](\frac{\alpha_m}{2}),\qquad z \ll 1.
\end{equation} 

This relationship between the redshift of the source and $\alpha_m$ had not been identified in any of the previous works of \citet{Espriu:2012be,Espriu:2014mwa} and \citet{Alfaro:2017sxd}, hence it represents a new and interesting finding in this line. This result implies that the $\Lambda$CDM model predicts large timing residuals for pulsars located approximately at the same direction of the (local) sources of gravitational waves. This effect could be useful for PTAs research and the measurement of Continuous Gravitational Waves through this method, a task that is expected to be accomplished in the next years \citep{Babak2016,Mingarelli2017,Nano2019_mm}: The angular separation between the sources of GWs and the monitored pulsars is strongly constrained in order to measure distinguishable differences in the timing residual. Therefore, if a PTA experiment seeks to measure a strong signal of gravitational radiation coming from single sources (e.g. mergers of supermassive black holes), the monitored pulsars must be located near to the angle $\alpha_m$ with respect to the sources. If that is not the case, the measurement of a powerful signal is very unlikely. In that sense, this prediction of the standard $\Lambda$CDM model represents an alternative way to improve the detection of GWs using PTA experiments by taking advantage of this local cosmological effect on the propagation of gravitational waves within our expanding Universe.
\section{Conclusions}
\label{sec:Conclusions}

In this manuscript we have extended the line of research developed in \cite{Bernabeu:2011if,Espriu:2012be,Espriu:2014mwa,Alfaro:2017sxd} on the action of cosmological parameters in the propagation of gravitational waves. This effect, an additional and different contribution besides the redshift in the frequency, is completely caused by the coordinate transformation between a frame centered onto the source of GWs and the comoving frame of the FLRW metric. The main result is that harmonic gravitational waves become anharmonic when a cosmological observer measures them, i.e. the Earth or PTAs. In particular, we have generalized the previous attempts to a case where all the cosmological contributions (e.g. Dark Energy, non-relativistic matter and radiation) are naturally included by the Hubble constant, in contrast to the previously developed idea of analyzing each of them separately, which seems to be wrong at first order in the perturbations. \\

We have shown that all the coordinate transformations involved are linear in terms of $H_0$, which turned out to be the parameter that governs the behavior of local cosmological fluctuations in the propagation of gravitational waves. From the expression \eqref{H0_approximation} an unexplored possibility of study the local behavior of $H_0$ opens, and besides, from \eqref{rs} we shown a direct relationship between the redshift of the source (for $z\ll 1$) and $\alpha_m$, in which the standard $\Lambda$CDM model predicts the observation of a strong signal in the time of arrival of electromagnetic emission of a monitored pulsar, when it is located at an angle close to $\alpha_m$ with respect to the source of continuous gravitational waves. \\

This local cosmological action appears to be very sensitive to PTA observations (basically due to the frequency spectrum of gravitational waves that PTAs could see). Although the process of measurement could be quite tough due to the experimental difficulties in the measurement of the astrophysical parameters involved (e.g. distances, redshift and timing residuals), some very privileged millisecond pulsars can be analyzed with an incredible accuracy \citep{Manchester:2017azn} and they could be useful in order to test the $\Lambda$CDM model within the near future. For this reason we believe it is worthwhile to perform a more detailed study of this phenomenon, for example, considering different sources or not--monochromatic GWs, in order to have a more realistic astrophysical analysis.

\section*{Acknowledgments}
The authors thank D. Espriu and L. Gabbanelli for many interesting conversations. M. Gamonal and J. Alfaro are partially supported by Fondecyt 1150390 and CONICYT-PIA-ACT14177.

\newpage
\begin{appendix}
\begin{section}{On the equations of standard Cosmology}
\label{sec:Appendix_Cosmology}

The components of the stress-energy tensor that appears in \eqref{linearized_EFE} can be found by considering a perfect fluid (i.e. a fluid that does not have viscosity and does not conduce heat), with energy density $\rho$ and isotropic pressure $p$, filling the whole Universe. The components of $T^{\mu\nu}$ take the following form,
\begin{equation}
T^{\mu\nu} = (\rho+ p) U^\mu U^\nu - pg^{\mu\nu},
\end{equation}
where $U^\mu$ are the components of the 4-velocity of the fluid. When this expression is inserted into the Einstein's Field equations we obtain the two \textit{Friedmann equations}: The first equation obtained from the 00 component of \eqref{EFE} and the second from the combination between the trace of the field equations and the first Friedmann equation, giving,
\begin{align}
\left(\frac{\dot{a}}{a}\right)^2 &= \frac{\kappa}{3} (\rho_i + \rho_\Lambda)\equiv H^2(T)	\label{Friedmann_1} \\
\left( \frac{\ddot{a}}{a}\right) &= \kappa \left( \frac{\rho_\Lambda}{3} - \frac{\rho_i}{6} - \frac{p_i}{2}  \right),	\label{Friedmann_2}
\end{align}
where $\rho_i$ and $p_i$ are the energy density and the isotropic pressure of the i-th fluid respectively and $H(T)$ is the Hubble parameter, which value at the present day, i.e. $H(T_0) \equiv H_0$, is known as the \textit{Hubble constant} $H_0$. In order to obtain the time evolution of the scale factor an equation of state must be provided. In \cite{Alfaro:2017sxd} was used $p=0$, which corresponds to the equation of state of non-relativistic dust. However, in this work we will use $p_i=\omega_i\rho_i$, with $\omega_i$ constant, in order to develop a general discussion of the phenomenon.  Using the Friedmann equations we can find that 
\begin{equation}
\label{rho_general}
\frac{\rho_i}{\rho_0} = \left( \frac{a(T)}{a_0} \right)^{-3(\omega_i+1)},
\end{equation}
where $\rho_0 = \rho(T_0)$ is the current energy density of the i-th fluid and $a_0 = a(T_0)$ is the current scale factor (usually taken as 1), and both are integration constants. Replacing this expression into \eqref{Friedmann_1} provides a solution of the scale factor in terms of the comoving time and the equation of state,

\begin{equation}
\label{a_general}
a(T) = 
\begin{cases}
a_0 \left(\frac{T}{T_0}\right)^{\frac{2}{3(\omega_i+1)}} & \textrm{if}\; \omega_i \neq -1\\
a_0\exp({\sqrt{\frac{\Lambda}{3}} (T-T_0)}) & \textrm{if} \; \omega_i = -1
\end{cases}
.
\end{equation}

In the $\omega_i\neq -1$ case, when we combine \eqref{rho_general} with \eqref{a_general} we can obtain the general form of the energy density of the i--th fluid,

\begin{equation}
\label{general_rho_i}
\rho_i = \begin{cases}
\frac{4}{3(\omega_i+1)^2 \kappa T^2} & \textrm{if}\; \omega_i \neq -1\\
\Lambda/\kappa & \textrm{if} \; \omega_i = -1
\end{cases}.
\end{equation}

Currently, the standard cosmological model is the $\Lambda$CDM: Includes a positive cosmological constant $\Lambda$ (which represents the so-called Dark Energy) and Cold Dark Matter (which is the union of baryonic matter and non-relativistic dark matter). In the $\Lambda$CDM model, when a global flat geometry is considered, we can use \eqref{rho_general} to write an effective energy density in terms of the scale factor and the currently evaluated energy densities,
\begin{align}
\rho_{\textrm{eff}} &= \rho_{\Lambda} + \rho_{d} + \rho_{r} \nonumber \\
&= \rho_{\Lambda} + \rho_{d0} \left[\frac{a_0}{a(T)}\right]^3 + \rho_{r0}\left[\frac{a_0}{a(T)}\right]^4,
\end{align}
where $\rho_{\Lambda} = \Lambda/\kappa$, $\rho_{d0}$ is the current density of non-relativistic matter (i.e. Cold Dark Matter and baryonic matter, $\omega_d = 0$) and $\rho_{r0}$ is the current radiation density ($\omega_r=1/3$). These expressions will be used in order to construct a spherically symmetric metric which reproduces the corresponding geometry of a FLRW metric for a perfect fluid with an arbitrary $\omega_i$. 
	\end{section}

\begin{section}{On the derivation of the SS$\omega_{\scriptsize{i}}$ metric}
	\label{sec:Appendix_SSw_metric}

	As we have to impose a spherically symmetric geometry we will have the transformation $r^2 \, \dd \Omega^2 \to a(T)^2 R^2 \, \dd \Omega^2$. Using the second rank tensor property of the metric tensor when we perform coordinate transformations,
	\begin{equation}
	\label{metric_second_tensor}
	g_{\mu' \nu'} = \pdv{X^\mu}{x^{\mu'}}\pdv{X^\nu}{x^{\nu'}} g_{\mu \nu},
	\end{equation}
	and the requirement that the new metric must be diagonal, we obtain the relation
	\begin{align}
	\label{SSw:grt=0}
	0 &=  \pdv{T}{t}\pdv{T}{r}g_{TT} + \pdv{R}{t} \pdv{R}{r} g_{RR}.
	\end{align}
	
	By computing the partial derivatives we obtain the expressions
	\begin{align}
	\pdv{R}{r} &= -\frac{1}{3} \frac{2r \pdv{T}{r} -3T (\omega_i+1)  }{a(T) (\omega_i+1) T} \label{SSw:dRdt}  \\
	\pdv{R}{t} &=  - \frac{2}{3} \frac{r \pdv{T}{t}}{a(T) (\omega_i+1) T},
	\label{SSw:dRdr}
	\end{align}
	and from \eqref{SSw:grt=0} we find that
	\begin{equation}
	\pdv{T}{r} = \frac{a(T)^2}{\pdv{T}{t}} \pdv{R}{t} \pdv{R}{r}.
	\end{equation}
	
	Thus, from the last equation, $\pdv{T}{r}$ becomes
	\begin{equation}
	\label{SSw:Tr}
	\pdv{T}{r} = \frac{6rT  (\omega_i+1)}{4r^2 - 9 (\omega_i+1)^2 T^2},
	\end{equation}
	and using \eqref{metric_second_tensor} we can obtain the components of the metric,
	\begin{align}
	g_{tt} &=- \Big( \pdv{T}{t} \Big)^2 \left[ \frac{9 (\omega_i+1)^2 T^2 - 4 r^2}{9(\omega_i+1)^2 T^2} \right] \label{SSw:gtt-funcionT} \\
	g_{rr} &= \frac{9(\omega_i+1)^2 T^2}{9(\omega_i+1)^2 T^2- 4r^2}. \label{SSw:grr-funcionT}
	\end{align}
	
	From \eqref{rho_general} we can write the SS$\omega_i$ metric as
	\begin{align}
	\dd s^2 = &-\frac{(\partial_t \rho_{i})^2}{3\kappa \rho_{i}^3(\omega_i+1)^2} \left[1-\frac{\kappa \rho_{i} r^2}{3} \right]\, \dd t^2 + \frac{\dd r^2}{1-\frac{\kappa \rho_{i} r^2}{3}} +r^2 \, \dd \Omega^2, \label{SSw:metric_close}
	\end{align}
	but, using \eqref{rho_general} and \eqref{SSw:Tr}, we get 
	\begin{equation}
	\label{SSw:rho_r}
	\pdv{\rho_{i}}{r} = \frac{(\omega_i+1)\kappa \rho_{i}^2 r}{1-\frac{\kappa \rho_{i}}{3} r^2}.
	\end{equation} 
	
	If we properly redefine  $\tilde{\rho}_{i} \equiv \kappa \rho_{i}$, the last expression becomes
	\begin{equation}
	\label{SSw:tilde{rho}_r}
	\pdv{\tilde{\rho}_{i}}{r} = \frac{ (\omega_i+1)\tilde{\rho}_{i}^2 r}{1-\frac{\tilde{\rho}_{i}}{3} r^2},
	\end{equation}
	but it can be noticed from \eqref{SSw:tilde{rho}_r} that we can form the expression
	\begin{equation}
	\label{pdv_r}
	\pdv{r} \left[ \frac{c+r^2 \tilde{\rho}_{i}}{\tilde{\rho}_{i}^{n}}  \right] = 0,
	\end{equation}
	where $c$ and $n$ are unknown constants that we suppose exist. Unfolding the last expression and using the linear independence of $r$, we obtain that the constants are
	\begin{equation}
	\label{SSw:c y n}
	c = \frac{6}{3\omega_i+1} \qquad n = \frac{3\omega_i+1}{3(\omega_i+1)}.
	\end{equation}
	
	Therefore, we can integrate \eqref{pdv_r} and write 
	\begin{equation}
	\label{SSw:tilde{rho}=C(t)}
	\frac{c+r^2 \tilde{\rho}_{i}}{\tilde{\rho}_{i}^{n}} = F(t),
	\end{equation}
	where $F(t)$ is a function of $t$. By a dimensional analysis, we note that in natural units $[\tilde{\rho}_{i}] = L^{-2}$ and therefore $[F(t)] = L^{2n}$. As there is no other parameter involved apart from $t$, and also as $[t] = L$ in natural units, then we set $F(t) = A t^{2n}$, with $A$ as a dimensionless arbitrary constant. For any fluid we can expect that at later stage it will be diluted homogeneously, which implies that for $t\to\infty$ the metric \eqref{SSw:metric_close} is almost flat. Then,
	\begin{equation}
	\lim_{t\to\infty  (\rho_{i}\to 0)}  \frac{(\partial_t \rho_{i})^2}{3\kappa \rho_{i}^3 (\omega_i+1)^2} = 1.
	\end{equation}
	
	On the other hand, \eqref{SSw:tilde{rho}=C(t)} can be written as
	\begin{equation}
	\label{SSw:rho y t}
	\frac{c+r^2 \kappa \rho_{i}}{(\kappa\rho_{i})^{n}} = At^{2n},
	\end{equation}
	but when we take the derivative with respect to $t$ and solving for $\partial_t \rho_i$, we obtain
	
	\begin{equation}
	\pdv{\rho_{i}}{t} = - \frac{2nA t^{2n-1} (\kappa \rho_{i})^n \rho_{i}}{\kappa \rho_{i} nr^2 - r^2 \kappa \rho_{i} + cn },
	\end{equation}
	and if we square, divide by $3\kappa \rho_{i}^3$ and replace the previous results, we can found the following equality 
	\begin{equation}
	\label{SSw:dt_krho}
	\frac{(\partial_t \rho_{i})^2}{3\kappa \rho_{i}^3(\omega_i+1)^2} = \frac{4n^2 A^{1/n} (\kappa r^2 \rho_{i} + c)^{\frac{2n-1}{n}}}{3(\omega_i+1)^2 [(n-1)\kappa r^2 \rho_{i} + cn]}.
	\end{equation}
	
	Computing the limit $\rho_{i} \to 0$ as the fluid dilutes at distant times, we can set $A$,
	\begin{equation}
	\lim_{t\to\infty  (\rho_{i}\to 0)}  \frac{(\partial_t \rho_{i})^2}{3\kappa \rho_{i}^3(\omega_i+1)^2} = \frac{4n^2 A^{1/n} c^{\frac{2n-1}{n}}}{3(\omega_i+1)^2 (cn)^2},
	\end{equation}
	and using that the metric is asymptotically flat, which implies that the previous limit is equal to one, we get the value of $A$,
	\begin{equation}
	A = c \left(\frac{3}{4}\right)^n (\omega_i+1)^{2n}.
	\end{equation}
	
	Finally, with the constant $A$ known, we can provide an exact expression for the the SS$\omega_i$ metric, which becomes
	
	\begin{align}
	\dd s^2 = &-\frac{\dd t^2}{\left(1-\dfrac{\kappa\rho_{i}r^2}{3} \right)\left( 1 + \dfrac{\kappa\rho_{i} r^2 (3\omega_i+1)}{6}  \right)^{\frac{1-3\omega_i}{1+3\omega_i}}} + \frac{\dd r^2}{1-\dfrac{\kappa\rho_{i}r^2}{3}} + r^2 \dd \Omega^2 ,
	\end{align}
	and from \eqref{SSw:rho y t} we can express the coordinate transformation between the SS$\omega_i$ and the FLRW frames in terms of $\rho_{i}$ y $\rho_{0}$,
	\begin{align}
	t &= \frac{\left[ c + R^2 (\kappa \rho_{0})^{\frac{2}{3(\omega_i+1)}}(\kappa \rho_i)^{\frac{3\omega_i+1}{3(\omega_i+1)}}  \right]^{\frac{1}{2n}}}{\left(A^{\frac{1}{2n}} \right)\sqrt{\kappa \rho_{i}}} \label{SSw:Final_transformations_A_appendix} \\
	r &= R\left( \frac{\rho_{0}}{\rho_{i}}  \right)^{\frac{1}{3(\omega_i+1)}}. \label{SSw:Final_transformations_B_appendix}
	\end{align}

\end{section}

\begin{section}{On the accuracy in the approximation of $H_0$}
	\label{sec:Appendix}

In order to simplify the computation, we can omit the geometrical prefactor that appears in \eqref{tau_GW_LCDM_full}, because it is common to every observation and is $H_0$--independent. Therefore, we define a reduced timing residual,
\begin{align}
\label{red_tau}
\tau_{\textrm{GW}}^{\textrm{red}} \equiv \int_{-1}^{0} \frac{1 + H_0 \left[ T_e + \frac{xL}{c} \right]   }{Z + xL\cos\alpha} \sin(\frac{\pi}{4} + \Theta(x,\alpha))\; \dd x \approx R_1 + \left(\frac{1+ \frac{H_0 Z}{c}}{Z}\right) \int_{-1}^{0} \sin(\frac{\pi}{4} + \Theta(x,\alpha))\; \dd x  ,
\end{align}
with $\abs{R_1} \leq \frac{L H_0}{c Z} \sim 10^{-31}$ s. Then, we take the reduced timing residual from \eqref{red_tau} and note that $R_1$ is given by

\begin{equation}
R_{1} = \int_{-1}^{0} \dd x  \sin \left( \Theta ( x, \alpha ) + \frac{\pi}{4}
\right) \left[ \frac{1+H_{0} \left[ \frac{Z_{e}}{c} + \frac{L}{c} x
	\right]}{Z_{e} +x L  \cos   \alpha} - \frac{1+H_{0} \frac{Z_{e}}{c}}{Z_{e}}
\right].
\end{equation}

Thus we can bound the value of $R_1$ by

\begin{align}
| R_{1} | &\le  \frac{L}{Z_{e}} \int_{-1}^{0}  \left| \sin \left(
\Theta ( x, \alpha ) + \frac{\pi}{4} \right) \right|  \times \left| \frac{H_{0}
	\left[ \frac{1}{c} x \right] -x  \frac{L}{Z_{e}^{2}}   \cos   \alpha -x 
	\frac{L}{Z_{e}}   \cos   \alpha H_{0} \frac{1}{c}}{\left( 1+x 
	\frac{L}{Z_{e}}   \cos   \alpha \right)} \right| \dd x \nonumber \\ 
&\le \frac{LH_0}{2Z c} + \order{\frac{L^2}{Z^3}} \sim 10^{-31}\,\textrm{s}.
\end{align}

Then we can reasonable neglect $R_1$ in the equation \eqref{red_tau}. Now we can express $\tau_{\textrm{GW}}^{\textrm{red}}$ in terms of the imaginary part of the complex exponential and write, since $\Theta ( x, \alpha )$ is quadratic in $x$:

\begin{align}
\label{tau_red_im_exp}
\tau_{\textrm{GW}}^{\textrm{red}} &=  \Im{  \int_{-1}^{0} \dd x \; e^{i \left( \Theta ( x, \alpha ) +
		\frac{\pi}{4} \right)} }= \Im{ B(\alpha) e^{i \left( \Theta ( x^{\ast} , \alpha ) + \frac{\pi}{4} \right)} } ,
\end{align}

\noindent where $B(\alpha)$ is defined as $
B(\alpha) \equiv \int_{-1}^{0} \dd x \; e^{i \lambda ( x-x^{\ast} )^{2}}$,  $x^*$ satisfies $\eval{\pdv*{\Theta(x,\alpha)}{x}}_{x=x^*} = 0$, thus

\begin{equation}
x^* = \frac{-c+c  \cos\alpha + \mathit{Z}_{e}
	\mathit{H_{0}}}{\left( \cos \alpha^{2} -2  \cos   \alpha
	\right) \mathit{H_{0}} L},
\end{equation}
\noindent and $\lambda$ is given by

\begin{equation}
\label{lambda_expansion} 
\lambda = \frac{1}{2} \pdv[2]{\Theta(x,\alpha)}{x} =
\frac{1}{2} \frac{\Omega H_{0} L^{2}}{c^{2}} ( \cos   \alpha^{2} -2  \cos  
\alpha ).
\end{equation}

The integral $B(\alpha)$ can be written in terms of the error function, giving

\begin{align}
B(\alpha) = \frac{\sqrt{2 \pi}}{4} ( 1+i ) \frac{1}{\sqrt{\lambda}} \left[ -
\erf \left( \frac{\sqrt{2}}{2} ( 1-i ) u^{\ast} \right) + \erf
\left( \frac{\sqrt{2}}{2} ( 1-i ) \left( \sqrt{\lambda} +u^{\ast} \right)
\right) \right],
\end{align}

where $u^* \equiv \sqrt{\lambda} x^*$. Using the asymptotic expansion of the error functions for $u^* \gg 1$  \citep[see][]{Abramowitz1972}, we can write

\begin{equation}
B(\alpha) \approx e^{-z_{1}^{2}}
\left( 1+ \frac{1}{2z_{1}} \right) \qquad z_1 \equiv \frac{\sqrt{2}}{2} ( 1-i ) u^{*}.
\end{equation}

Inserting the last expression into \eqref{tau_red_im_exp}, $\tau_{\textrm{GW}}^{\textrm{red}}$ becomes

\begin{equation}
\tau_{\textrm{GW}}^{\textrm{red}} \approx  \sin \left( C+ \frac{\pi}{4} \right) + \frac{1}{2 | u^{\ast} |} \sin  C,
\end{equation}
where $C=H_0 Z^2 \Omega/2c^2$. From this expression we can see that the maximum of $\tau_{\textrm{GW}}^{\textrm{red}}$ clearly happens for $u^*\to 0$. This condition  implies, from \eqref{lambda_expansion}, that the angle corresponding to the maximum absolute value of $\tau_{\textrm{GW}}$ satisfies $x^*=0$, or, rearranging the terms, the approximation formula \eqref{H0_approximation}. In order to justify the validity of the asymptotic expansion, we can explore around $u^*=0$, finding that for a variation in the angle $\Delta \alpha$, then $u^* \sim i \sqrt{\frac{Z\Omega}{c}} \Delta \alpha \sim 10^{4} \Delta \alpha $. Thus, the expansion is well defined for $\Delta \alpha\gg 10^{-4}$.

\end{section}

\newpage

\begin{section}{Table of pulsars of the ATNF catalog}
	\label{sec:Appendix_pulsars}
	
	\begin{table}[h!]
		\centering
		\begin{adjustbox}{width=0.2\textheight}
			\begin{tabular}{|c|ccc|}
				\hline
				Pulsar Name&$\theta$&$\phi$& $L_i$\\
				\hline
				J0324+5239 & $168.5^\circ$ & $-31.68^\circ$ & 2.56 kpc \\ 
				J0325+67 & $145^\circ$ & $-1.22^\circ$ & 1.51 kpc \\ 
				J0329+1654 & $130.31^\circ$ & $18.68^\circ$ & 1.05 kpc \\ 
				J0332+5434 & $150.35^\circ$ & $-8.64^\circ$ & 1.54 kpc \\ 
				J0332+79 & $169.99^\circ$ & $-30.04^\circ$ & 1.30 kpc \\
				\hline
				J2007+2722 & $78.23^\circ$ & $2.09^\circ$ & 2.15 kpc \\ 
				J2007+3120 & $68.86^\circ$ & $-4.67^\circ$ & 2.10 kpc \\ 
				J2008+2513 & $76.89^\circ$ & $0.96^\circ$ & 10.3 kpc \\ 
				J2009+3326 & $87.86^\circ$ & $8.38^\circ$ & 1.83 kpc \\ 
				J2010-1323  & $86.86^\circ$ & $7.54^\circ$ & 2.06 kpc \\
				\hline
				J1848-1150 & $35.26^\circ$ & $1.4^\circ$ & 12.3 kpc \\ 
				J1848+12 & $36.72^\circ$ & $2.23^\circ$ & 8.23 kpc \\ 
				J1848-1243 & $44.99^\circ$ & $6.34^\circ$ & 2.17 kpc \\ 
				J1848-1414 & $46.69^\circ$ & $7.29^\circ$ & 1.22 kpc \\ 
				J1848-1952  & $32.54^\circ$ & $-0.33^\circ$ & 5.45 kpc \\
				\hline
				J1826-1256 & $21.33^\circ$ & $0.26^\circ$ & 4.94 kpc \\ 
				J1826-1334 & $14.6^\circ$ & $-3.42^\circ$ & 5.94 kpc \\ 
				J1826-1419 & $53.34^\circ$ & $15.61^\circ$ & 0.91 kpc \\ 
				J1826-1526 & $29.76^\circ$ & $4.25^\circ$ & 10.3 kpc \\ 
				J1827-0750  & $29.16^\circ$ & $3.99^\circ$ & 3.50 kpc \\
				\hline
				J1946+14 & $66.86^\circ$ & $2.55^\circ$ & 7.47 kpc \\ 
				J1946+1805 & $44.86^\circ$ & $-10.55^\circ$ & 3.94 kpc \\ 
				J1946+2052 & $61.1^\circ$ & $-1.17^\circ$ & 7.27 kpc \\ 
				J1946+2244 & $50^\circ$ & $-7.74^\circ$ & 1.51 kpc \\ 
				J1946+24  & $52.5^\circ$ & $-6.58^\circ$ & 1.59 kpc \\
				\hline
				J1946+24 & $30.81^\circ$ & $3.73^\circ$ & 3.35 kpc \\ 
				J1831-1329 & $30.57^\circ$ & $3.45^\circ$ & 4.68 kpc \\ 
				J1831-1423 & $27.04^\circ$ & $1.75^\circ$ & 2.49 kpc \\ 
				J1832+0029 & $25.64^\circ$ & $0.96^\circ$ & 6.30 kpc \\ 
				J1832-0644  & $25.17^\circ$ & $0.76^\circ$ & 8.29 kpc \\
				\hline
				J2155-3118 & $108.64^\circ$ & $6.85^\circ$ & 1.88 kpc \\ 
				J2155-5641 & $89.66^\circ$ & $-22.81^\circ$ & 2.82 kpc \\ 
				J2156+2618 & $87.69^\circ$ & $-26.28^\circ$ & 1.80 kpc \\ 
				J2157+4017 & $106.65^\circ$ & $2.95^\circ$ & 3.00 kpc \\ 
				J2203+50  & $107.15^\circ$ & $3.64^\circ$ & 3.01 kpc \\
				\hline
				J1840-0809  & $30.28^\circ$ & $1.02^\circ$ & 6.97 kpc \\ 
				J1840-0815  & $34.56^\circ$ & $3.34^\circ$ & 5.04 kpc \\ 
				J1840-0840 & $29.08^\circ$ & $0.58^\circ$ & 8.58 kpc \\ 
				J1840-1122  & $35.43^\circ$ & $3.85^\circ$ & 4.33 kpc \\ 
				J1840-1207  & $28.35^\circ$ & $0.17^\circ$ & 3.71 kpc \\
				\hline
				J1828-2119  & $31.25^\circ$ & $4.36^\circ$ & 1.04 kpc \\ 
				J1829+0000  & $24.81^\circ$ & $1.07^\circ$ & 10.4 kpc \\ 
				J1829-0734 & $23.27^\circ$ & $0.3^\circ$ & 5.20 kpc \\ 
				J1829-1011  & $23.11^\circ$ & $0.26^\circ$ & 0.81 kpc \\ 
				J1829-1751  & $21.59^\circ$ & $-0.6^\circ$ & 4.69 kpc \\
				\hline
				J1848-0511  & $32.76^\circ$ & $0.09^\circ$ & 5.63 kpc \\ 
				J1848-0601  & $32.41^\circ$ & $0.07^\circ$ & 6.71 kpc \\ 
				J1848+0604 & $33.25^\circ$ & $0.35^\circ$ & 4.05 kpc \\ 
				J1848+0647  & $32.37^\circ$ & $-0.04^\circ$ & 6.5 kpc \\ 
				J1848+0826  & $34.02^\circ$ & $0.96^\circ$ & 3.39 kpc \\
				\hline
				J1843-0050  & $29.57^\circ$ & $0.12^\circ$ & 6.03 kpc \\ 
				J1843-0137  & $29.52^\circ$ & $0.07^\circ$ & 5.70 kpc \\ 
				J1843-0211 & $29.4^\circ$ & $0.24^\circ$ & 5.26 kpc \\ 
				J1843-0355  & $29.34^\circ$ & $0.04^\circ$ & 5.97 kpc \\ 
				J1843-0408   & $28.79^\circ$ & $-0.19^\circ$ & 5.45 kpc \\
				\hline
			\end{tabular}
		\end{adjustbox}
		\caption{\label{tab:pulsars_real} List of randomly distributed pulsars averaged for an hypothetical source. The galactic longitude is denoted by 6$\theta$ and the galactic latitude by $\phi$. More information about the pulsars can be found \href{http://www.atnf.csiro.au/research/pulsar/psrcat/}{here}.}
	\end{table}

	\end{section}
	\end{appendix}


\begin{thebibliography}{45}
	
	%\expandafter\ifx\csname natexlab\endcsname\relax\def\natexlab#1{#1}\fi
	\expandafter\ifx\csname bibnamefont\endcsname\relax
	\def\bibnamefont#1{#1}\fi
	\expandafter\ifx\csname bibfnamefont\endcsname\relax
	\def\bibfnamefont#1{#1}\fi
	\expandafter\ifx\csname citenamefont\endcsname\relax
	\def\citenamefont#1{#1}\fi
	\expandafter\ifx\csname url\endcsname\relax
	\def\url#1{\texttt{#1}}\fi
	\expandafter\ifx\csname urlprefix\endcsname\relax\def\urlprefix{URL }\fi
	\providecommand{\bibinfo}[2]{#2}
	\providecommand{\eprint}[2][]{\url{#2}}
	
	\bibitem[{\citenamefont{{Perlmutter} et~al.}(1997)\citenamefont{{Perlmutter},
			{Gabi}, {Goldhaber}, {Goobar}, {Groom}, {Hook}, {Kim}, {Kim}, {Lee}, {Pain}
			et~al.}}]{Perlmutter1996}
	\bibinfo{author}{\bibfnamefont{S.}~\bibnamefont{{Perlmutter}}},
	\bibinfo{author}{\bibfnamefont{S.}~\bibnamefont{{Gabi}}},
	\bibinfo{author}{\bibfnamefont{G.}~\bibnamefont{{Goldhaber}}},
	\bibinfo{author}{\bibfnamefont{A.}~\bibnamefont{{Goobar}}},
	\bibinfo{author}{\bibfnamefont{D.~E.} \bibnamefont{{Groom}}},
	\bibinfo{author}{\bibfnamefont{I.~M.} \bibnamefont{{Hook}}},
	\bibinfo{author}{\bibfnamefont{A.~G.} \bibnamefont{{Kim}}},
	\bibinfo{author}{\bibfnamefont{M.~Y.} \bibnamefont{{Kim}}},
	\bibinfo{author}{\bibfnamefont{J.~C.} \bibnamefont{{Lee}}},
	\bibinfo{author}{\bibfnamefont{R.}~\bibnamefont{{Pain}}},
	\bibnamefont{et~al.}, \bibinfo{journal}{\apj} \textbf{\bibinfo{volume}{483}},
	\bibinfo{pages}{565} (\bibinfo{year}{1997}), \eprint{astro-ph/9608192}.
	
	\bibitem[{\citenamefont{{Riess} et~al.}(1998)\citenamefont{{Riess},
			{Filippenko}, {Challis}, {Clocchiatti}, {Diercks}, {Garnavich}, {Gilliland},
			{Hogan}, {Jha}, {Kirshner} et~al.}}]{Riess1998}
	\bibinfo{author}{\bibfnamefont{A.~G.} \bibnamefont{{Riess}}},
	\bibinfo{author}{\bibfnamefont{A.~V.} \bibnamefont{{Filippenko}}},
	\bibinfo{author}{\bibfnamefont{P.}~\bibnamefont{{Challis}}},
	\bibinfo{author}{\bibfnamefont{A.}~\bibnamefont{{Clocchiatti}}},
	\bibinfo{author}{\bibfnamefont{A.}~\bibnamefont{{Diercks}}},
	\bibinfo{author}{\bibfnamefont{P.~M.} \bibnamefont{{Garnavich}}},
	\bibinfo{author}{\bibfnamefont{R.~L.} \bibnamefont{{Gilliland}}},
	\bibinfo{author}{\bibfnamefont{C.~J.} \bibnamefont{{Hogan}}},
	\bibinfo{author}{\bibfnamefont{S.}~\bibnamefont{{Jha}}},
	\bibinfo{author}{\bibfnamefont{R.~P.} \bibnamefont{{Kirshner}}},
	\bibnamefont{et~al.}, \bibinfo{journal}{\aj} \textbf{\bibinfo{volume}{116}},
	\bibinfo{pages}{1009} (\bibinfo{year}{1998}), \eprint{astro-ph/9805201}.
	
	\bibitem[{\citenamefont{{Planck Collaboration}
			et~al.}(2018)\citenamefont{{Planck Collaboration}, {Aghanim}, {Akrami},
			{Ashdown}, {Aumont}, {Baccigalupi}, {Ballardini}, {Banday}, {Barreiro},
			{Bartolo} et~al.}}]{Planck2018}
	\bibinfo{author}{\bibnamefont{{Planck Collaboration}}},
	\bibinfo{author}{\bibfnamefont{N.}~\bibnamefont{{Aghanim}}},
	\bibinfo{author}{\bibfnamefont{Y.}~\bibnamefont{{Akrami}}},
	\bibinfo{author}{\bibfnamefont{M.}~\bibnamefont{{Ashdown}}},
	\bibinfo{author}{\bibfnamefont{J.}~\bibnamefont{{Aumont}}},
	\bibinfo{author}{\bibfnamefont{C.}~\bibnamefont{{Baccigalupi}}},
	\bibinfo{author}{\bibfnamefont{M.}~\bibnamefont{{Ballardini}}},
	\bibinfo{author}{\bibfnamefont{A.~J.} \bibnamefont{{Banday}}},
	\bibinfo{author}{\bibfnamefont{R.~B.} \bibnamefont{{Barreiro}}},
	\bibinfo{author}{\bibfnamefont{N.}~\bibnamefont{{Bartolo}}},
	\bibnamefont{et~al.}, \bibinfo{journal}{arXiv e-prints}
	\bibinfo{eid}{arXiv:1807.06209} (\bibinfo{year}{2018}), \eprint{1807.06209}.
	
	\bibitem[{\citenamefont{{Riess} et~al.}(2018)\citenamefont{{Riess},
			{Casertano}, {Yuan}, {Macri}, {Bucciarelli}, {Lattanzi}, {MacKenty},
			{Bowers}, {Zheng}, {Filippenko} et~al.}}]{Riess2018}
	\bibinfo{author}{\bibfnamefont{A.~G.} \bibnamefont{{Riess}}},
	\bibinfo{author}{\bibfnamefont{S.}~\bibnamefont{{Casertano}}},
	\bibinfo{author}{\bibfnamefont{W.}~\bibnamefont{{Yuan}}},
	\bibinfo{author}{\bibfnamefont{L.}~\bibnamefont{{Macri}}},
	\bibinfo{author}{\bibfnamefont{B.}~\bibnamefont{{Bucciarelli}}},
	\bibinfo{author}{\bibfnamefont{M.~G.} \bibnamefont{{Lattanzi}}},
	\bibinfo{author}{\bibfnamefont{J.~W.} \bibnamefont{{MacKenty}}},
	\bibinfo{author}{\bibfnamefont{J.~B.} \bibnamefont{{Bowers}}},
	\bibinfo{author}{\bibfnamefont{W.}~\bibnamefont{{Zheng}}},
	\bibinfo{author}{\bibfnamefont{A.~V.} \bibnamefont{{Filippenko}}},
	\bibnamefont{et~al.}, \bibinfo{journal}{\apj} \textbf{\bibinfo{volume}{861}},
	\bibinfo{eid}{126} (\bibinfo{year}{2018}), \eprint{1804.10655}.
	
	\bibitem[{\citenamefont{{Hotokezaka} et~al.}(2019)\citenamefont{{Hotokezaka},
			{Nakar}, {Gottlieb}, {Nissanke}, {Masuda}, {Hallinan}, {Mooley}, and
			{Deller}}}]{LIGO2019_GW}
	\bibinfo{author}{\bibfnamefont{K.}~\bibnamefont{{Hotokezaka}}},
	\bibinfo{author}{\bibfnamefont{E.}~\bibnamefont{{Nakar}}},
	\bibinfo{author}{\bibfnamefont{O.}~\bibnamefont{{Gottlieb}}},
	\bibinfo{author}{\bibfnamefont{S.}~\bibnamefont{{Nissanke}}},
	\bibinfo{author}{\bibfnamefont{K.}~\bibnamefont{{Masuda}}},
	\bibinfo{author}{\bibfnamefont{G.}~\bibnamefont{{Hallinan}}},
	\bibinfo{author}{\bibfnamefont{K.~P.} \bibnamefont{{Mooley}}},
	\bibnamefont{and} \bibinfo{author}{\bibfnamefont{A.~T.}
		\bibnamefont{{Deller}}}, \bibinfo{journal}{Nature Astronomy} p.
	\bibinfo{pages}{385} (\bibinfo{year}{2019}).
	
	\bibitem[{\citenamefont{{Wong} et~al.}(2019)\citenamefont{{Wong}, {Suyu},
			{Chen}, {Rusu}, {Millon}, {Sluse}, {Bonvin}, {Fassnacht}, {Taubenberger},
			{Auger} et~al.}}]{Holicow2019}
	\bibinfo{author}{\bibfnamefont{K.~C.} \bibnamefont{{Wong}}},
	\bibinfo{author}{\bibfnamefont{S.~H.} \bibnamefont{{Suyu}}},
	\bibinfo{author}{\bibfnamefont{G.~C.~F.} \bibnamefont{{Chen}}},
	\bibinfo{author}{\bibfnamefont{C.~E.} \bibnamefont{{Rusu}}},
	\bibinfo{author}{\bibfnamefont{M.}~\bibnamefont{{Millon}}},
	\bibinfo{author}{\bibfnamefont{D.}~\bibnamefont{{Sluse}}},
	\bibinfo{author}{\bibfnamefont{V.}~\bibnamefont{{Bonvin}}},
	\bibinfo{author}{\bibfnamefont{C.~D.} \bibnamefont{{Fassnacht}}},
	\bibinfo{author}{\bibfnamefont{S.}~\bibnamefont{{Taubenberger}}},
	\bibinfo{author}{\bibfnamefont{M.~W.} \bibnamefont{{Auger}}},
	\bibnamefont{et~al.} (\bibinfo{year}{2019}), \bibinfo{note}{preprint at
		arXiv:1907.04869}.
	
	\bibitem[{\citenamefont{Einstein}(1916)}]{Einstein1916}
	\bibinfo{author}{\bibfnamefont{A.}~\bibnamefont{Einstein}},
	\bibinfo{journal}{Annalen Phys.} \textbf{\bibinfo{volume}{49}},
	\bibinfo{pages}{769} (\bibinfo{year}{1916}).
	
	\bibitem[{\citenamefont{{Odderskov} et~al.}(2014)\citenamefont{{Odderskov},
			{Hannestad}, and {Haugb{\o}lle}}}]{Odderskov2014}
	\bibinfo{author}{\bibfnamefont{I.}~\bibnamefont{{Odderskov}}},
	\bibinfo{author}{\bibfnamefont{S.}~\bibnamefont{{Hannestad}}},
	\bibnamefont{and}
	\bibinfo{author}{\bibfnamefont{T.}~\bibnamefont{{Haugb{\o}lle}}},
	\bibinfo{journal}{Journal of Cosmology and Astro-Particle Physics}
	\textbf{\bibinfo{volume}{2014}}, \bibinfo{eid}{028} (\bibinfo{year}{2014}),
	\eprint{1407.7364}.
	
	\bibitem[{\citenamefont{{Ko} and {Tang}}(2016)}]{Ko2016}
	\bibinfo{author}{\bibfnamefont{P.}~\bibnamefont{{Ko}}} \bibnamefont{and}
	\bibinfo{author}{\bibfnamefont{Y.}~\bibnamefont{{Tang}}},
	\bibinfo{journal}{Physics Letters B} \textbf{\bibinfo{volume}{762}},
	\bibinfo{pages}{462} (\bibinfo{year}{2016}), \eprint{1608.01083}.
	
	\bibitem[{\citenamefont{{Freedman}}(2017)}]{Freedman2017}
	\bibinfo{author}{\bibfnamefont{W.~L.} \bibnamefont{{Freedman}}},
	\bibinfo{journal}{Nature Astronomy} \textbf{\bibinfo{volume}{1}},
	\bibinfo{eid}{0169} (\bibinfo{year}{2017}), \eprint{1706.02739}.
	
	\bibitem[{\citenamefont{{Bringmann} et~al.}(2018)\citenamefont{{Bringmann},
			{Kahlhoefer}, {Schmidt-Hoberg}, and {Walia}}}]{Bringmann2018}
	\bibinfo{author}{\bibfnamefont{T.}~\bibnamefont{{Bringmann}}},
	\bibinfo{author}{\bibfnamefont{F.}~\bibnamefont{{Kahlhoefer}}},
	\bibinfo{author}{\bibfnamefont{K.}~\bibnamefont{{Schmidt-Hoberg}}},
	\bibnamefont{and} \bibinfo{author}{\bibfnamefont{P.}~\bibnamefont{{Walia}}},
	\bibinfo{journal}{\prd} \textbf{\bibinfo{volume}{98}}, \bibinfo{eid}{023543}
	(\bibinfo{year}{2018}), \eprint{1803.03644}.
	
	\bibitem[{\citenamefont{{Camarena} and {Marra}}(2018)}]{Camarena2018}
	\bibinfo{author}{\bibfnamefont{D.}~\bibnamefont{{Camarena}}} \bibnamefont{and}
	\bibinfo{author}{\bibfnamefont{V.}~\bibnamefont{{Marra}}},
	\bibinfo{journal}{\prd} \textbf{\bibinfo{volume}{98}}, \bibinfo{eid}{023537}
	(\bibinfo{year}{2018}), \eprint{1805.09900}.
	
	\bibitem[{\citenamefont{{M{\"o}rtsell} and {Dhawan}}(2018)}]{Mortsell2018}
	\bibinfo{author}{\bibfnamefont{E.}~\bibnamefont{{M{\"o}rtsell}}}
	\bibnamefont{and} \bibinfo{author}{\bibfnamefont{S.}~\bibnamefont{{Dhawan}}},
	\bibinfo{journal}{Journal of Cosmology and Astro-Particle Physics}
	\textbf{\bibinfo{volume}{2018}}, \bibinfo{eid}{025} (\bibinfo{year}{2018}),
	\eprint{1801.07260}.
	
	\bibitem[{\citenamefont{{Di Valentino} et~al.}(2018)\citenamefont{{Di
				Valentino}, {Linder}, and {Melchiorri}}}]{DiValentino2018}
	\bibinfo{author}{\bibfnamefont{E.}~\bibnamefont{{Di Valentino}}},
	\bibinfo{author}{\bibfnamefont{E.~V.} \bibnamefont{{Linder}}},
	\bibnamefont{and} \bibinfo{author}{\bibfnamefont{A.~r.}
		\bibnamefont{{Melchiorri}}}, \bibinfo{journal}{\prd}
	\textbf{\bibinfo{volume}{97}}, \bibinfo{eid}{043528} (\bibinfo{year}{2018}),
	\eprint{1710.02153}.
	
	\bibitem[{\citenamefont{Feeney et~al.}(2019)\citenamefont{Feeney, Peiris,
			Williamson, Nissanke, Mortlock, Alsing, and Scolnic}}]{Feeney2019}
	\bibinfo{author}{\bibfnamefont{S.~M.} \bibnamefont{Feeney}},
	\bibinfo{author}{\bibfnamefont{H.~V.} \bibnamefont{Peiris}},
	\bibinfo{author}{\bibfnamefont{A.~R.} \bibnamefont{Williamson}},
	\bibinfo{author}{\bibfnamefont{S.~M.} \bibnamefont{Nissanke}},
	\bibinfo{author}{\bibfnamefont{D.~J.} \bibnamefont{Mortlock}},
	\bibinfo{author}{\bibfnamefont{J.}~\bibnamefont{Alsing}}, \bibnamefont{and}
	\bibinfo{author}{\bibfnamefont{D.}~\bibnamefont{Scolnic}},
	\bibinfo{journal}{Phys. Rev. Lett.} \textbf{\bibinfo{volume}{122}},
	\bibinfo{pages}{061105} (\bibinfo{year}{2019}),
	\urlprefix\url{https://link.aps.org/doi/10.1103/PhysRevLett.122.061105}.
	
	\bibitem[{\citenamefont{{Abbott} et~al.}(2016)\citenamefont{{Abbott}, {Abbott},
			{Abbott}, {Abernathy}, {Acernese}, {Ackley}, {Adams}, {Adams}, {Addesso},
			{Adhikari} et~al.}}]{Ligo2016}
	\bibinfo{author}{\bibfnamefont{B.~P.} \bibnamefont{{Abbott}}},
	\bibinfo{author}{\bibfnamefont{R.}~\bibnamefont{{Abbott}}},
	\bibinfo{author}{\bibfnamefont{T.~D.} \bibnamefont{{Abbott}}},
	\bibinfo{author}{\bibfnamefont{M.~R.} \bibnamefont{{Abernathy}}},
	\bibinfo{author}{\bibfnamefont{F.}~\bibnamefont{{Acernese}}},
	\bibinfo{author}{\bibfnamefont{K.}~\bibnamefont{{Ackley}}},
	\bibinfo{author}{\bibfnamefont{C.}~\bibnamefont{{Adams}}},
	\bibinfo{author}{\bibfnamefont{T.}~\bibnamefont{{Adams}}},
	\bibinfo{author}{\bibfnamefont{P.}~\bibnamefont{{Addesso}}},
	\bibinfo{author}{\bibfnamefont{R.~X.} \bibnamefont{{Adhikari}}},
	\bibnamefont{et~al.}, \bibinfo{journal}{\prl} \textbf{\bibinfo{volume}{116}},
	\bibinfo{eid}{061102} (\bibinfo{year}{2016}), \eprint{1602.03837}.
	
	\bibitem[{\citenamefont{{LIGO Collaboration} et~al.}(2017)\citenamefont{{LIGO
				Collaboration}, Abbott et~al.}}]{Ligo2017}
	\bibinfo{author}{\bibnamefont{{LIGO Collaboration}}},
	\bibinfo{author}{\bibfnamefont{B.~P.} \bibnamefont{Abbott}},
	\bibnamefont{et~al.} (\bibinfo{collaboration}{LIGO Scientific, Virgo, 1M2H,
		Dark Energy Camera GW-E, DES, DLT40, Las Cumbres Observatory, VINROUGE,
		MASTER}), \bibinfo{journal}{Nature} \textbf{\bibinfo{volume}{551}},
	\bibinfo{pages}{85} (\bibinfo{year}{2017}), \eprint{1710.05835}.
	
	\bibitem[{\citenamefont{{Barke} et~al.}(2015)\citenamefont{{Barke}, {Wang},
			{Esteban Delgado}, {Tr{\"o}bs}, {Heinzel}, and {Danzmann}}}]{Barke2015}
	\bibinfo{author}{\bibfnamefont{S.}~\bibnamefont{{Barke}}},
	\bibinfo{author}{\bibfnamefont{Y.}~\bibnamefont{{Wang}}},
	\bibinfo{author}{\bibfnamefont{J.~J.} \bibnamefont{{Esteban Delgado}}},
	\bibinfo{author}{\bibfnamefont{M.}~\bibnamefont{{Tr{\"o}bs}}},
	\bibinfo{author}{\bibfnamefont{G.}~\bibnamefont{{Heinzel}}},
	\bibnamefont{and}
	\bibinfo{author}{\bibfnamefont{K.}~\bibnamefont{{Danzmann}}},
	\bibinfo{journal}{Class. Quant. Grav.} \textbf{\bibinfo{volume}{32}},
	\bibinfo{eid}{095004} (\bibinfo{year}{2015}), \eprint{1411.1260}.
	
	\bibitem[{\citenamefont{Hobbs and Dai}(2017)}]{Hobbs2017}
	\bibinfo{author}{\bibfnamefont{G.}~\bibnamefont{Hobbs}} \bibnamefont{and}
	\bibinfo{author}{\bibfnamefont{S.}~\bibnamefont{Dai}},
	\bibinfo{journal}{Natl. Sci. Rev.} \textbf{\bibinfo{volume}{4}},
	\bibinfo{pages}{707} (\bibinfo{year}{2017}).
	
	\bibitem[{\citenamefont{{Cordes} et~al.}(2019)\citenamefont{{Cordes},
			{McLaughlin}, and {Nanograv Collaboration}}}]{Nano2019_fp}
	\bibinfo{author}{\bibfnamefont{J.}~\bibnamefont{{Cordes}}},
	\bibinfo{author}{\bibfnamefont{M.~A.} \bibnamefont{{McLaughlin}}},
	\bibnamefont{and} \bibinfo{author}{\bibnamefont{{Nanograv Collaboration}}},
	\bibinfo{journal}{Bull. Amer. Astron. Soc.} \textbf{\bibinfo{volume}{51}},
	\bibinfo{eid}{447} (\bibinfo{year}{2019}).
	
	\bibitem[{\citenamefont{{Burke-Spolaor}
			et~al.}(2019)\citenamefont{{Burke-Spolaor}, {Taylor}, {Charisi}, {Dolch},
			{Hazboun}, {Holgado}, {Kelley}, {Lazio}, {Madison}, {McMann}
			et~al.}}]{Burke2019}
	\bibinfo{author}{\bibfnamefont{S.}~\bibnamefont{{Burke-Spolaor}}},
	\bibinfo{author}{\bibfnamefont{S.~R.} \bibnamefont{{Taylor}}},
	\bibinfo{author}{\bibfnamefont{M.}~\bibnamefont{{Charisi}}},
	\bibinfo{author}{\bibfnamefont{T.}~\bibnamefont{{Dolch}}},
	\bibinfo{author}{\bibfnamefont{J.~S.} \bibnamefont{{Hazboun}}},
	\bibinfo{author}{\bibfnamefont{A.~M.} \bibnamefont{{Holgado}}},
	\bibinfo{author}{\bibfnamefont{L.~Z.} \bibnamefont{{Kelley}}},
	\bibinfo{author}{\bibfnamefont{T.~J.~W.} \bibnamefont{{Lazio}}},
	\bibinfo{author}{\bibfnamefont{D.~R.} \bibnamefont{{Madison}}},
	\bibinfo{author}{\bibfnamefont{N.}~\bibnamefont{{McMann}}},
	\bibnamefont{et~al.}, \bibinfo{journal}{Astron. Astrophys. Rev.}
	\textbf{\bibinfo{volume}{27}}, \bibinfo{eid}{5} (\bibinfo{year}{2019}).
	
	\bibitem[{\citenamefont{{McLaughlin}}(2013)}]{NanoGrav2013}
	\bibinfo{author}{\bibfnamefont{M.~A.} \bibnamefont{{McLaughlin}}},
	\bibinfo{journal}{Class. Quant. Grav.} \textbf{\bibinfo{volume}{30}},
	\bibinfo{eid}{224008} (\bibinfo{year}{2013}).
	
	\bibitem[{\citenamefont{{Hobbs}}(2013)}]{Parkes2013}
	\bibinfo{author}{\bibfnamefont{G.}~\bibnamefont{{Hobbs}}},
	\bibinfo{journal}{Class. Quant. Grav.} \textbf{\bibinfo{volume}{30}},
	\bibinfo{eid}{224007} (\bibinfo{year}{2013}).
	
	\bibitem[{\citenamefont{{Ferdman} et~al.}(2010)\citenamefont{{Ferdman}, {van
				Haasteren}, {Bassa}, {Burgay}, {Cognard}, {Corongiu}, {D'Amico}, {Desvignes},
			{Hessels}, {Janssen} et~al.}}]{EPTA2010}
	\bibinfo{author}{\bibfnamefont{R.~D.} \bibnamefont{{Ferdman}}},
	\bibinfo{author}{\bibfnamefont{R.}~\bibnamefont{{van Haasteren}}},
	\bibinfo{author}{\bibfnamefont{C.~G.} \bibnamefont{{Bassa}}},
	\bibinfo{author}{\bibfnamefont{M.}~\bibnamefont{{Burgay}}},
	\bibinfo{author}{\bibfnamefont{I.}~\bibnamefont{{Cognard}}},
	\bibinfo{author}{\bibfnamefont{A.}~\bibnamefont{{Corongiu}}},
	\bibinfo{author}{\bibfnamefont{N.}~\bibnamefont{{D'Amico}}},
	\bibinfo{author}{\bibfnamefont{G.}~\bibnamefont{{Desvignes}}},
	\bibinfo{author}{\bibfnamefont{J.~W.~T.} \bibnamefont{{Hessels}}},
	\bibinfo{author}{\bibfnamefont{G.~H.} \bibnamefont{{Janssen}}},
	\bibnamefont{et~al.}, \bibinfo{journal}{Class. Quant. Grav.}
	\textbf{\bibinfo{volume}{27}}, \bibinfo{eid}{084014} (\bibinfo{year}{2010}).
	
	\bibitem[{\citenamefont{{Hobbs} et~al.}(2010)\citenamefont{{Hobbs},
			{Archibald}, {Arzoumanian}, {Backer}, {Bailes}, {Bhat}, {Burgay},
			{Burke-Spolaor}, {Champion}, {Cognard} et~al.}}]{Hobbs2010}
	\bibinfo{author}{\bibfnamefont{G.}~\bibnamefont{{Hobbs}}},
	\bibinfo{author}{\bibfnamefont{A.}~\bibnamefont{{Archibald}}},
	\bibinfo{author}{\bibfnamefont{Z.}~\bibnamefont{{Arzoumanian}}},
	\bibinfo{author}{\bibfnamefont{D.}~\bibnamefont{{Backer}}},
	\bibinfo{author}{\bibfnamefont{M.}~\bibnamefont{{Bailes}}},
	\bibinfo{author}{\bibfnamefont{N.~D.~R.} \bibnamefont{{Bhat}}},
	\bibinfo{author}{\bibfnamefont{M.}~\bibnamefont{{Burgay}}},
	\bibinfo{author}{\bibfnamefont{S.}~\bibnamefont{{Burke-Spolaor}}},
	\bibinfo{author}{\bibfnamefont{D.}~\bibnamefont{{Champion}}},
	\bibinfo{author}{\bibfnamefont{I.}~\bibnamefont{{Cognard}}},
	\bibnamefont{et~al.}, \bibinfo{journal}{Classical and Quantum Gravity}
	\textbf{\bibinfo{volume}{27}}, \bibinfo{eid}{084013} (\bibinfo{year}{2010}),
	\eprint{0911.5206}.
	
	\bibitem[{\citenamefont{{Lentati} et~al.}(2015)\citenamefont{{Lentati},
			{Taylor}, {Mingarelli}, {Sesana}, {Sanidas}, {Vecchio}, {Caballero}, {Lee},
			{van Haasteren}, {Babak} et~al.}}]{EPTA_BG}
	\bibinfo{author}{\bibfnamefont{L.}~\bibnamefont{{Lentati}}},
	\bibinfo{author}{\bibfnamefont{S.~R.} \bibnamefont{{Taylor}}},
	\bibinfo{author}{\bibfnamefont{C.~M.~F.} \bibnamefont{{Mingarelli}}},
	\bibinfo{author}{\bibfnamefont{A.}~\bibnamefont{{Sesana}}},
	\bibinfo{author}{\bibfnamefont{S.~A.} \bibnamefont{{Sanidas}}},
	\bibinfo{author}{\bibfnamefont{A.}~\bibnamefont{{Vecchio}}},
	\bibinfo{author}{\bibfnamefont{R.~N.} \bibnamefont{{Caballero}}},
	\bibinfo{author}{\bibfnamefont{K.~J.} \bibnamefont{{Lee}}},
	\bibinfo{author}{\bibfnamefont{R.}~\bibnamefont{{van Haasteren}}},
	\bibinfo{author}{\bibfnamefont{S.}~\bibnamefont{{Babak}}},
	\bibnamefont{et~al.}, \bibinfo{journal}{Mon. Notices Royal Astron. Soc.}
	\textbf{\bibinfo{volume}{453}}, \bibinfo{pages}{2576} (\bibinfo{year}{2015}).
	
	\bibitem[{\citenamefont{{Verbiest} et~al.}(2016)\citenamefont{{Verbiest},
			{Lentati}, {Hobbs}, {van Haasteren}, {Demorest}, {Janssen}, {Wang},
			{Desvignes}, {Caballero}, {Keith} et~al.}}]{IPTA2016}
	\bibinfo{author}{\bibfnamefont{J.~P.~W.} \bibnamefont{{Verbiest}}},
	\bibinfo{author}{\bibfnamefont{L.}~\bibnamefont{{Lentati}}},
	\bibinfo{author}{\bibfnamefont{G.}~\bibnamefont{{Hobbs}}},
	\bibinfo{author}{\bibfnamefont{R.}~\bibnamefont{{van Haasteren}}},
	\bibinfo{author}{\bibfnamefont{P.~B.} \bibnamefont{{Demorest}}},
	\bibinfo{author}{\bibfnamefont{G.~H.} \bibnamefont{{Janssen}}},
	\bibinfo{author}{\bibfnamefont{J.~B.} \bibnamefont{{Wang}}},
	\bibinfo{author}{\bibfnamefont{G.}~\bibnamefont{{Desvignes}}},
	\bibinfo{author}{\bibfnamefont{R.~N.} \bibnamefont{{Caballero}}},
	\bibinfo{author}{\bibfnamefont{M.~J.} \bibnamefont{{Keith}}},
	\bibnamefont{et~al.}, \bibinfo{journal}{\mnras}
	\textbf{\bibinfo{volume}{458}}, \bibinfo{pages}{1267} (\bibinfo{year}{2016}),
	\eprint{1602.03640}.
	
	\bibitem[{\citenamefont{{Arzoumanian} et~al.}(2018)\citenamefont{{Arzoumanian},
			{Baker}, {Brazier}, {Burke-Spolaor}, {Chamberlin}, {Chatterjee}, {Christy},
			{Cordes}, {Cornish}, {Crawford} et~al.}}]{NANOGrav}
	\bibinfo{author}{\bibfnamefont{Z.}~\bibnamefont{{Arzoumanian}}},
	\bibinfo{author}{\bibfnamefont{P.~T.} \bibnamefont{{Baker}}},
	\bibinfo{author}{\bibfnamefont{A.}~\bibnamefont{{Brazier}}},
	\bibinfo{author}{\bibfnamefont{S.}~\bibnamefont{{Burke-Spolaor}}},
	\bibinfo{author}{\bibfnamefont{S.~J.} \bibnamefont{{Chamberlin}}},
	\bibinfo{author}{\bibfnamefont{S.}~\bibnamefont{{Chatterjee}}},
	\bibinfo{author}{\bibfnamefont{B.}~\bibnamefont{{Christy}}},
	\bibinfo{author}{\bibfnamefont{J.~M.} \bibnamefont{{Cordes}}},
	\bibinfo{author}{\bibfnamefont{N.~J.} \bibnamefont{{Cornish}}},
	\bibinfo{author}{\bibfnamefont{F.}~\bibnamefont{{Crawford}}},
	\bibnamefont{et~al.}, \bibinfo{journal}{\apj} \textbf{\bibinfo{volume}{859}},
	\bibinfo{eid}{47} (\bibinfo{year}{2018}), \eprint{1801.02617}.
	
	\bibitem[{\citenamefont{{Babak} et~al.}(2016)\citenamefont{{Babak}, {Petiteau},
			{Sesana}, {Brem}, {Rosado}, {Taylor}, {Lassus}, {Hessels}, {Bassa}, {Burgay}
			et~al.}}]{Babak2016}
	\bibinfo{author}{\bibfnamefont{S.}~\bibnamefont{{Babak}}},
	\bibinfo{author}{\bibfnamefont{A.}~\bibnamefont{{Petiteau}}},
	\bibinfo{author}{\bibfnamefont{A.}~\bibnamefont{{Sesana}}},
	\bibinfo{author}{\bibfnamefont{P.}~\bibnamefont{{Brem}}},
	\bibinfo{author}{\bibfnamefont{P.~A.} \bibnamefont{{Rosado}}},
	\bibinfo{author}{\bibfnamefont{S.~R.} \bibnamefont{{Taylor}}},
	\bibinfo{author}{\bibfnamefont{A.}~\bibnamefont{{Lassus}}},
	\bibinfo{author}{\bibfnamefont{J.~W.~T.} \bibnamefont{{Hessels}}},
	\bibinfo{author}{\bibfnamefont{C.~G.} \bibnamefont{{Bassa}}},
	\bibinfo{author}{\bibfnamefont{M.}~\bibnamefont{{Burgay}}},
	\bibnamefont{et~al.}, \bibinfo{journal}{\mnras}
	\textbf{\bibinfo{volume}{455}}, \bibinfo{pages}{1665} (\bibinfo{year}{2016}),
	\eprint{1509.02165}.
	
	\bibitem[{\citenamefont{{Mingarelli} et~al.}(2017)\citenamefont{{Mingarelli},
			{Lazio}, {Sesana}, {Greene}, {Ellis}, {Ma}, {Croft}, {Burke-Spolaor}, and
			{Taylor}}}]{Mingarelli2017}
	\bibinfo{author}{\bibfnamefont{C.~M.~F.} \bibnamefont{{Mingarelli}}},
	\bibinfo{author}{\bibfnamefont{T.~J.~W.} \bibnamefont{{Lazio}}},
	\bibinfo{author}{\bibfnamefont{A.}~\bibnamefont{{Sesana}}},
	\bibinfo{author}{\bibfnamefont{J.~E.} \bibnamefont{{Greene}}},
	\bibinfo{author}{\bibfnamefont{J.~A.} \bibnamefont{{Ellis}}},
	\bibinfo{author}{\bibfnamefont{C.-P.} \bibnamefont{{Ma}}},
	\bibinfo{author}{\bibfnamefont{S.}~\bibnamefont{{Croft}}},
	\bibinfo{author}{\bibfnamefont{S.}~\bibnamefont{{Burke-Spolaor}}},
	\bibnamefont{and} \bibinfo{author}{\bibfnamefont{S.~R.}
		\bibnamefont{{Taylor}}}, \bibinfo{journal}{Nature Astronomy}
	\textbf{\bibinfo{volume}{1}}, \bibinfo{pages}{886} (\bibinfo{year}{2017}).
	
	\bibitem[{\citenamefont{{Kelley} et~al.}(2019)\citenamefont{{Kelley},
			{Charisi}, {Burke-Spolaor}, {Simon}, {Blecha}, {Bogdanovic}, {Colpi},
			{Comerford}, {D'Orazio}, {Dotti} et~al.}}]{Nano2019_mm}
	\bibinfo{author}{\bibfnamefont{L.}~\bibnamefont{{Kelley}}},
	\bibinfo{author}{\bibfnamefont{M.}~\bibnamefont{{Charisi}}},
	\bibinfo{author}{\bibfnamefont{S.}~\bibnamefont{{Burke-Spolaor}}},
	\bibinfo{author}{\bibfnamefont{J.}~\bibnamefont{{Simon}}},
	\bibinfo{author}{\bibfnamefont{L.}~\bibnamefont{{Blecha}}},
	\bibinfo{author}{\bibfnamefont{T.}~\bibnamefont{{Bogdanovic}}},
	\bibinfo{author}{\bibfnamefont{M.}~\bibnamefont{{Colpi}}},
	\bibinfo{author}{\bibfnamefont{J.}~\bibnamefont{{Comerford}}},
	\bibinfo{author}{\bibfnamefont{D.}~\bibnamefont{{D'Orazio}}},
	\bibinfo{author}{\bibfnamefont{M.}~\bibnamefont{{Dotti}}},
	\bibnamefont{et~al.}, \bibinfo{journal}{Bull. Amer. Astron. Soc.}
	\textbf{\bibinfo{volume}{51}}, \bibinfo{eid}{490} (\bibinfo{year}{2019}).
	
	\bibitem[{\citenamefont{{Perera} et~al.}(2019)\citenamefont{{Perera},
			{DeCesar}, {Demorest}, {Kerr}, {Lentati}, {Nice}, {Os{\l}owski}, {Ransom},
			{Keith}, {Arzoumanian} et~al.}}]{IPTA_2DR}
	\bibinfo{author}{\bibfnamefont{B.~B.~P.} \bibnamefont{{Perera}}},
	\bibinfo{author}{\bibfnamefont{M.~E.} \bibnamefont{{DeCesar}}},
	\bibinfo{author}{\bibfnamefont{P.~B.} \bibnamefont{{Demorest}}},
	\bibinfo{author}{\bibfnamefont{M.}~\bibnamefont{{Kerr}}},
	\bibinfo{author}{\bibfnamefont{L.}~\bibnamefont{{Lentati}}},
	\bibinfo{author}{\bibfnamefont{D.~J.} \bibnamefont{{Nice}}},
	\bibinfo{author}{\bibfnamefont{S.}~\bibnamefont{{Os{\l}owski}}},
	\bibinfo{author}{\bibfnamefont{S.~M.} \bibnamefont{{Ransom}}},
	\bibinfo{author}{\bibfnamefont{M.~J.} \bibnamefont{{Keith}}},
	\bibinfo{author}{\bibfnamefont{Z.}~\bibnamefont{{Arzoumanian}}},
	\bibnamefont{et~al.}, \bibinfo{journal}{\mnras}
	\textbf{\bibinfo{volume}{490}}, \bibinfo{pages}{4666} (\bibinfo{year}{2019}),
	\eprint{1909.04534}.
	
	\bibitem[{\citenamefont{{Bernabeu} et~al.}(2011)\citenamefont{{Bernabeu},
			{Espriu}, and {Puigdom{\`e}nech}}}]{Bernabeu:2011if}
	\bibinfo{author}{\bibfnamefont{J.}~\bibnamefont{{Bernabeu}}},
	\bibinfo{author}{\bibfnamefont{D.}~\bibnamefont{{Espriu}}}, \bibnamefont{and}
	\bibinfo{author}{\bibfnamefont{D.}~\bibnamefont{{Puigdom{\`e}nech}}},
	\bibinfo{journal}{\prd} \textbf{\bibinfo{volume}{84}}, \bibinfo{eid}{063523}
	(\bibinfo{year}{2011}), \eprint{1106.4511}.
	
	\bibitem[{\citenamefont{{Espriu} and {Puigdom{\`e}nech}}(2013)}]{Espriu:2012be}
	\bibinfo{author}{\bibfnamefont{D.}~\bibnamefont{{Espriu}}} \bibnamefont{and}
	\bibinfo{author}{\bibfnamefont{D.}~\bibnamefont{{Puigdom{\`e}nech}}},
	\bibinfo{journal}{\apj} \textbf{\bibinfo{volume}{764}}, \bibinfo{eid}{163}
	(\bibinfo{year}{2013}), \eprint{1209.3724}.
	
	\bibitem[{\citenamefont{{Espriu}}(2014)}]{Espriu:2014mwa}
	\bibinfo{author}{\bibfnamefont{D.}~\bibnamefont{{Espriu}}}, in
	\emph{\bibinfo{booktitle}{American Institute of Physics Conference Series}}
	(\bibinfo{year}{2014}), vol. \bibinfo{volume}{1606} of
	\emph{\bibinfo{series}{American Institute of Physics Conference Series}}, pp.
	\bibinfo{pages}{86--98}, \eprint{1401.7925}.
	
	\bibitem[{\citenamefont{{Alfaro} et~al.}(2019)\citenamefont{{Alfaro}, {Espriu},
			and {Gabbanelli}}}]{Alfaro:2017sxd}
	\bibinfo{author}{\bibfnamefont{J.}~\bibnamefont{{Alfaro}}},
	\bibinfo{author}{\bibfnamefont{D.}~\bibnamefont{{Espriu}}}, \bibnamefont{and}
	\bibinfo{author}{\bibfnamefont{L.}~\bibnamefont{{Gabbanelli}}},
	\bibinfo{journal}{Class. Quant. Grav.} \textbf{\bibinfo{volume}{36}},
	\bibinfo{eid}{025006} (\bibinfo{year}{2019}), \eprint{1711.08315}.
	
	\bibitem[{\citenamefont{{Einstein}}(1916)}]{Einstein1916_GW}
	\bibinfo{author}{\bibfnamefont{A.}~\bibnamefont{{Einstein}}},
	\bibinfo{journal}{Sitzungsberichte der K{\"o}niglich Preu{\ss}ischen Akademie
		der Wissenschaften (Berlin), Seite 688-696.}  (\bibinfo{year}{1916}).
	
	\bibitem[{\citenamefont{Cheng}(2010)}]{cheng}
	\bibinfo{author}{\bibfnamefont{T.~P.} \bibnamefont{Cheng}},
	\emph{\bibinfo{title}{Relativity, Gravitation and Cosmology, a basic
			introduction}} (\bibinfo{publisher}{Oxford University Press},
	\bibinfo{year}{2010}).
	
	\bibitem[{\citenamefont{{Cervantes-Cota} and {Smoot}}(2011)}]{Cervantes2011}
	\bibinfo{author}{\bibfnamefont{J.~L.} \bibnamefont{{Cervantes-Cota}}}
	\bibnamefont{and} \bibinfo{author}{\bibfnamefont{G.}~\bibnamefont{{Smoot}}},
	in \emph{\bibinfo{booktitle}{American Institute of Physics Conference
			Series}}, edited by \bibinfo{editor}{\bibfnamefont{L.~A.}
		\bibnamefont{{Ure{\~n}a-L{\'o}pez}}},
	\bibinfo{editor}{\bibfnamefont{H.}~\bibnamefont{{Aurelio
				Morales-T{\'e}cotl}}},
	\bibinfo{editor}{\bibfnamefont{R.}~\bibnamefont{{Linares-Romero}}},
	\bibinfo{editor}{\bibfnamefont{E.}~\bibnamefont{{Santos-Rodr{\'\i}guez}}},
	\bibnamefont{and}
	\bibinfo{editor}{\bibfnamefont{S.}~\bibnamefont{{Estrada-Jim{\'e}nez}}}
	(\bibinfo{year}{2011}), vol. \bibinfo{volume}{1396} of
	\emph{\bibinfo{series}{American Institute of Physics Conference Series}}, pp.
	\bibinfo{pages}{28--52}, \eprint{1107.1789}.
	
	\bibitem[{\citenamefont{{Finn}}(2009)}]{Finn2009}
	\bibinfo{author}{\bibfnamefont{L.~S.} \bibnamefont{{Finn}}},
	\bibinfo{journal}{\prd} \textbf{\bibinfo{volume}{79}}, \bibinfo{eid}{022002}
	(\bibinfo{year}{2009}), \eprint{0810.4529}.
	
	\bibitem[{\citenamefont{{Deng} and {Finn}}(2011)}]{Deng2011}
	\bibinfo{author}{\bibfnamefont{X.}~\bibnamefont{{Deng}}} \bibnamefont{and}
	\bibinfo{author}{\bibfnamefont{L.~S.} \bibnamefont{{Finn}}},
	\bibinfo{journal}{\mnras} \textbf{\bibinfo{volume}{414}}, \bibinfo{pages}{50}
	(\bibinfo{year}{2011}), \eprint{1008.0320}.
	
	\bibitem[{\citenamefont{{Manchester} et~al.}(2005)\citenamefont{{Manchester},
			{Hobbs}, {Teoh}, and {Hobbs}}}]{Catalogo}
	\bibinfo{author}{\bibfnamefont{R.~N.} \bibnamefont{{Manchester}}},
	\bibinfo{author}{\bibfnamefont{G.~B.} \bibnamefont{{Hobbs}}},
	\bibinfo{author}{\bibfnamefont{A.}~\bibnamefont{{Teoh}}}, \bibnamefont{and}
	\bibinfo{author}{\bibfnamefont{M.}~\bibnamefont{{Hobbs}}},
	\bibinfo{journal}{\aj} \textbf{\bibinfo{volume}{129}}, \bibinfo{pages}{1993}
	(\bibinfo{year}{2005}), \eprint{astro-ph/0412641}.
	
	\bibitem[{\citenamefont{Ryden}(2003)}]{Ryden}
	\bibinfo{author}{\bibfnamefont{B.}~\bibnamefont{Ryden}},
	\emph{\bibinfo{title}{Introduction to Cosmology}}
	(\bibinfo{publisher}{Addison-Wesley}, \bibinfo{address}{San Francisco},
	\bibinfo{year}{2003}).
	
	\bibitem[{\citenamefont{{Manchester}}(2017)}]{Manchester:2017azn}
	\bibinfo{author}{\bibfnamefont{R.~N.} \bibnamefont{{Manchester}}},
	\bibinfo{journal}{Journal of Astrophysics and Astronomy}
	\textbf{\bibinfo{volume}{38}}, \bibinfo{eid}{42} (\bibinfo{year}{2017}),
	\eprint{1709.09434}.
	
	\bibitem[{\citenamefont{Abramowitz and Stegun}(1972)}]{Abramowitz1972}
	\bibinfo{author}{\bibfnamefont{M.}~\bibnamefont{Abramowitz}} \bibnamefont{and}
	\bibinfo{author}{\bibfnamefont{I.}~\bibnamefont{Stegun}},
	\emph{\bibinfo{title}{Handbook of mathematical functions with formulas,
			graphs, and mathematical tables}} (\bibinfo{publisher}{Dept. of Commerce,
		National Bureau of Standards.}, \bibinfo{year}{1972}).
	
\end{thebibliography}
\end{document}